\documentclass[aps,pre,reprint,longbibliography,nobalancelastpage]{revtex4-2}

\usepackage{amsmath,amssymb,amsthm,amsbsy}
\usepackage{graphicx}
\usepackage{xcolor}
\usepackage{mathrsfs}
\usepackage{stmaryrd}
\usepackage{hyperref}
\usepackage{enumitem}
\usepackage{accents}
\usepackage{tikz}
\usetikzlibrary{tikzmark}

\newcommand{\tikzarc}[1]{%
\tikzmarknode{a}{#1}
\begin{tikzpicture}[overlay,remember picture]
\draw ([yshift=.5pt]a.north west) to[bend left=20] ([yshift=.5pt]a.north east);
\end{tikzpicture}%
}

\definecolor{linkcolor}{HTML}{223096}
\hypersetup{colorlinks,allcolors=linkcolor}
\bibpunct{\textcolor{linkcolor}{[}}{\textcolor{linkcolor}{]}}{\textcolor{linkcolor}{,}}{n}{}{;}
\renewcommand{\eqref}[1]{\hyperref[#1]{(\ref*{#1})}}

\usetikzlibrary{decorations.pathreplacing,arrows}

\renewcommand{\vec}[1]{\boldsymbol{#1}}

\newcommand{\figref}[2]{[Fig.~\hyperref[#1]{\ref*{#1}(#2)}]}
\newcommand{\figrefi}[2]{[Fig.~\hyperref[#1]{\ref*{#1}(#2)}, inset]}
\newcommand{\textfigref}[2]{Fig.~\hyperref[#1]{\ref*{#1}(#2)}}
\newcommand{\textfigureref}[2]{Figure~\hyperref[#1]{\ref*{#1}(#2)}}

\newcommand{\figrefp}[2]{\hyperref[#1]{\ref*{#1}(#2)}}

\renewcommand{\leq}{\leqslant}
\renewcommand{\geq}{\geqslant}

\newtheoremstyle{note}{3pt}{3pt}{}{}{\itshape}{.}{.5em}{}
\theoremstyle{note}
\newtheorem*{lemma}{Lemma}

\begin{document}

\title{Geometry of T1 transitions in epithelia}
\author{Pierre A. Haas}
\email{haas@pks.mpg.de}
\affiliation{Max Planck Institute for the Physics of Complex Systems, N\"othnitzer Stra\ss e 38, 01187 Dresden, Germany}
\affiliation{\smash{Max Planck Institute of Molecular Cell Biology and Genetics, Pfotenhauerstra\ss e 108, 01307 Dresden, Germany}}
\affiliation{Center for Systems Biology Dresden, Pfotenhauerstra\ss e 108, 01307 Dresden, Germany}
\date{\today}%
\begin{abstract}
The flows of tissues of epithelial cells often involve T1 transitions. These neighbour exchanges are irreversible rearrangements crossing an energy barrier. Here, by an exact geometric construction, I determine this energy barrier for general, isolated T1 transitions dominated by line tensions. I~show how deviations from regular cell packing reduce this energy barrier, but find that line tension fluctuations increase it on average. By another exact construction, I prove that the nonlinear tensions in vertex models of tissues also resist T1 transitions. My results thus form the basis for coarse-grained understanding of cell neighbour exchanges for continuum descriptions of epithelia.
\end{abstract}

\maketitle
\begin{figure}[b]
\includegraphics{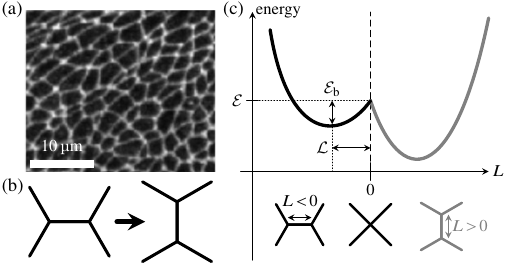}
\caption{T1 transitions. (a) Example of a two-dimensional epithelial tissue, the \emph{Drosophila} wing disc epithelium, taken from Ref.~\cite{dye21}. Lines are cell boundaries. Scale bar: $10\,\text{\textmu m}$. (b)~Schematic of a T1 transition: before the T1 transition, the ``top'' and ``bottom'' cells are neighbours; after the transition, the ``left'' and ``right'' cells are neighbours. (c)~Generic energy landscape of a T1 transition, plotted against the T1 transition coordinate $L$, the signed length of the central edge. The energy landscape has a cusp at the T1 transition point $L=0$. The energy minimum for $L<0$ is at $L=-\mathcal{L}$, where $\mathcal{L}>0$. This and the energy $\mathcal{E}$ at the T1 transition define the energy barrier $\mathcal{E}_\text{b}$. The energy landscape for $L>0$ is similar. Shape insets: T1 transition and definition of $L$.\label{fig1}}
\end{figure}

The deformations of epithelial tissues~\figref{fig1}{a} in development are often associated with cell intercalations~\cite{walck14,rauzi20} during which initially neighbouring cells are separated by changes of the topology of the network of cell-cell contacts~\figref{fig1}{b}. Cell intercalations are involved in particular in the processes of convergence and extension~\cite{wolpert00,wallingford02} that pervade morphogenesis; an example is germband extension in \emph{Drosophila}~\cite{irvine94,bertet04,kong17,brauns24}.

In the language of two-dimensional foams~\cite{cantat}, these cell neighbour exchanges are T1 transitions. Mechanically, they are irreversible rearrangements of the tissue which involve the crossing of a barrier $\mathcal{E}_\text{b}$ in the energy landscape~\figref{fig1}{c} near the T1 transition, which has a cuspy shape~\cite{bi14,krajnc18,popovic21}. The vanishing of this energy barrier is associated with tissue fluidisation~\cite{bi15}, which has important biological roles in development and disease~\cite{hannezo22,lenne22}, for example in zebrafish morphogenesis~\cite{mongera18,petridou19,petridou21}.

Biological tissues are not of course two-dimensional, and their three-dimensional nature gives rise to complex cell neighbour relationships~\cite{gomezgalvez18,gomez21,lou23}, in which for example apical and basal cell surfaces have different cell-cell contact networks. In turn, this gives rise to a richer zoo of cell neighbour exchanges~\cite{sarkar24}. Still, this two-dimensional picture has proven useful to elucidate the physical principles governing the interplay between the tensions in cell-cell junctions, their fluctuations, and the resulting active T1 transitions and fluidisation of tissues~\cite{bi14,bi15,krajnc18,staddon19,popovic21,yan19,krajnc20,krajnc21,duclut21,duclut22,jain23,jain24,staddon25,demarzio25}. In particular, recent work has shown how mechanochemical feedbacks can generate such active T1 transitions~\cite{sknepnek23}.

There is however a more geometric component to the mechanics of T1 transitions, too: Epithelial tissues often feature large amounts of cell shape variability, as exemplified by the \emph{Drosophila} wing disc epithelium~\cite{dye21} shown in \textfigref{fig1}{a}. This disorder takes the geometry of a T1 transition far away from the regular T1 transition sketched in \textfigref{fig1}{b}, but its effect has remained unexplored.

Here, I show how this cell-scale disorder affects the energy barrier to isolated T1 transitions: I start from an exact geometric construction of a general T1 transition in the case in which the mechanics are dominated by homogeneous line tensions. I go on to extend these exact results to inhomogeneous and nonlinear line tensions.

\begin{figure}
\includegraphics{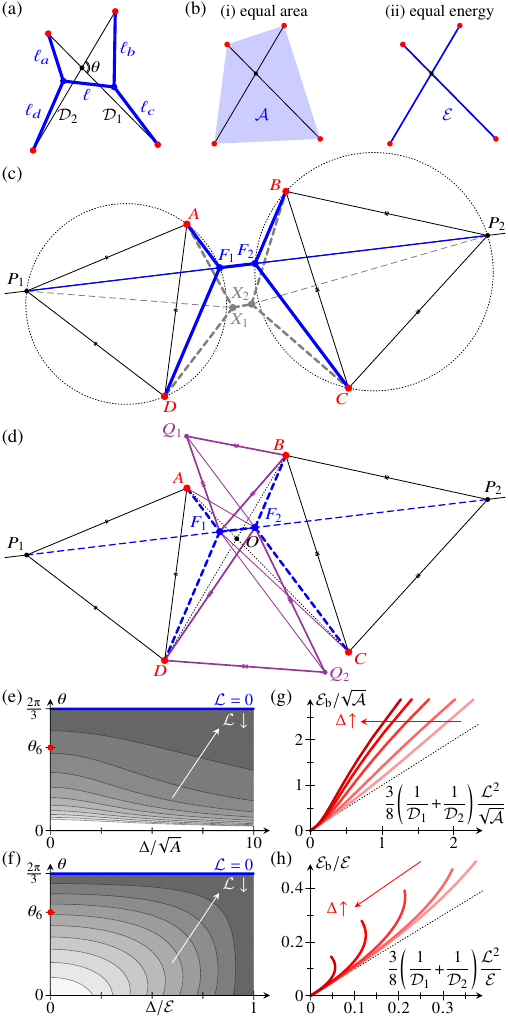}
\end{figure}

\begin{figure}
\vspace{-8pt}
\caption{Geometry of an isolated T1 transition dominated by (homogeneous) line tensions. (a) Plot of an isolated T1 transition: The outer vertices (red) are fixed as the central edge formed by the inner vertices (blue) undergoes a T1 transition. These outer vertices define a quadrilateral with diagonals $\mathcal{D}_1,\mathcal{D}_2$ that are at an angle $\theta$ to each other. (b)~Ensembles of isolated T1 transitions: (i) In the equal-area ensemble, the area $\mathcal{A}$ of the quadrilateral is fixed. (ii) In the equal-energy ensemble, the energy $\mathcal{E}$ of the transition configuration is held constant. (c)~Geometric construction of the positions $F_1,F_2$ of the inner vertices minimising the line-tension energy for fixed outer vertices $A,B,C,D$. See text for further explanation. (d)~Calculation of the energy barrier $\mathcal{E}_\text{b}$ and the distance~$\mathcal{L}$ from the T1 transition based on this geometric construction. See text for further explanation. (e) Contour plot of $\mathcal{L}$ against $\theta$ and the asymmetry $\Delta=|\mathcal{D}_1-\mathcal{D}_2|$ in the equal-area ensemble (i). The red mark corresponds to a regular hexagonal lattice $\theta=\theta_6$, $\Delta=0$. On the (blue) line $\theta=2\pi/3$, $\mathcal{L}=0$. (f)~Analogous plot for the equal-energy ensemble (ii). (g) Plot of the energy barrier $\mathcal{E}_\text{b}$ against $\mathcal{L}^2$, for the equal-area ensemble (i) and for different values of $\Delta$. The dotted line shows the asymptotic relation $\mathcal{E}_\text{b}/\mathcal{L}^2\sim 3(\mathcal{D}_1+\mathcal{D}_2)/(8\mathcal{D}_1\mathcal{D}_2)$. (h)~Analogous plot for the equal-energy ensemble (ii). \label{fig2}}
\begin{tikzpicture}[>=stealth]
\draw[<-] (0,0) -- (8.6,0);
\end{tikzpicture}
\end{figure}

An isolated T1 transition is defined by six cell vertices: four ``outer'' vertices held fixed and two ``inner'' vertices that undergo the T1 transition~\figref{fig2}{a}. The four outer vertices define a (convex) quadrilateral with diagonals $\mathcal{D}_1,\mathcal{D}_2$ making an angle $\theta$ with each other. In a regular hexagonal lattice of cells, $\theta=\theta_6\equiv \tan^{-1}{4\sqrt{3}}$ and $\mathcal{D}_1=\mathcal{D}_2$. This regular case underpins mean-field arguments for tissue fluidisation~\cite{bi15,gomezgalvez18,staddon25}. To analyse the general geometry, it will be useful to introduce two geometric ensembles~\figref{fig2}{b}: (i) an equal-area ensemble in which the area $\mathcal{A}$ of the quadrilateral defined by the four outer vertices is held constant, and (ii) an equal-energy ensemble in which the energy $\mathcal{E}$ at the point of the T1 transition~\figref{fig1}{c} is held constant instead. 

I begin by assuming that the mechanics are dominated by homogeneous line tensions. I denote by $\ell_a,\ell_b,\ell_c,\ell_d,\ell$ the lengths of the cell edges, shown in \textfigref{fig2}{a}, so that the energy is $E=\ell_a+\ell_b+\ell_c+\ell_d+\ell$. Labelling the four outer vertices $A,B,C,D$ as shown in \textfigref{fig2}{c}, I erect equilateral triangles $ADP_1$ and $BCP_2$ outwards on line segments $[AD]$ and $[BC]$. Line $P_1P_2$ intersects their circumcircles again at $F_1$ and $F_2$, respectively. I claim that $F_1,F_2$ are the positions of the inner vertices that minimise $E$. The proof relies on the following rephrasing of Pompeiu's theorem of classical Euclidean geometry~\cite{andreescu}:
\begin{lemma}
Let $ABC$ be an equilateral triangle. For any point $P$, $|AP|+|BP|\geq|CP|$, with equality if and only if $P$ lies on the minor arc $\tikzarc{AB}$ of the circumcircle of $ABC$.
\end{lemma}
\noindent Let $X_1,X_2$ be points in the plane, so that $\ell_a=|AX_1|$, $\ell_b=|BX_2|$, $\ell_c=|CX_2|$, $\ell_d=|DX_1|$, $\ell=|X_1X_2|$, as shown in \textfigref{fig2}{c}. Thus, by the above lemma,
\begin{align}
E&=\bigl(|AX_1|+|DX_1|\bigr)+|X_1X_2|+\bigl(|BX_2|+|CX_2|\bigr)\nonumber\\
&\geq |P_1X_1|+|X_1X_2|+|P_2X_2|\geq |P_1P_2|,\label{eq:E}
\end{align}
in which the final inequality follows from the triangle inequality. Equality holds throughout if and only if the following three conditions hold: (1) $X_1$ lies on the minor arc $\tikzarc{AD}$ of the circumcircle of triangle $P_1AD$, (2) $X_2$ lies on the minor arc \smash{$\tikzarc{BC}$} of the circumcircle of triangle $P_2BC$, and (3) points $P_1,X_1,X_2,P_2$ lie on a line in this order. These conditions imply $X_1=F_1$, $X_2=F_2$. An equivalent construction was obtained in Ref.~\cite{du87} as the solution of the Steiner minimal tree problem~\cite{gilbert68} on four vertices; I add that Ref.~\cite{du87} analyses carefully degenerate cases that I have neglected~\cite{[{Reference~\cite{du87} and }][{ show that the existence of the points $F_1,F_2$ and hence of the ``Steiner tree'' in \textfigref{fig2}{d} is equivalent with $\angle DAP_2\leq 120^\circ$, $\angle P_2DA\leq 120^\circ$, $\angle P_1BC\leq 120^\circ$, $\angle BCP_1\leq 120^\circ$, and $\theta\leq 120^\circ$. In this work, I shall assume tacitly that the first four conditions hold. This is of course a strong mathematical assumption, but not perhaps a strong physical assumption on the geometry of cell vertices surrounding a T1 transition edge.}]weng94}.

This construction requires $P_1,F_1,F_2,P_2$ to be in this order. Let $O$ denote the intersection of the diagonals $AC$, $BD$, as in \textfigref{fig2}{d}. Angles $\angle AF_1P_1$, $\angle ADP_1$ intercept the same chord of the circumcircle of the cyclic quadrilateral $AF_1DP_1$ so ${\angle AF_1P_1\!=\!\angle ADP_1\!=\!60^\circ}$ since triangle $ADP_1$ is equilateral. Similarly, ${\angle CF_2P_2\!=\!60^\circ}$, whence $AF_1\parallel CF_2$. Analogously, ${BF_2\parallel DF_1}$. This shows that ${F_1=F_2=F}$ if and only if ${F=O}$, in which case ${\theta=\angle AF_1D=180^\circ-\angle DP_1A=120^\circ}$. It is clear that $P_1,F_1,F_2,P_2$ are in this order for small $\theta$ and that $F_1,F_2$ vary continuously on line $P_1P_2$ as $\theta$ increases. Hence $P_1,F_1,F_2,P_2$ are in this order and the minimum $E=|P_1P_2|$ in Eq.~\eqref{eq:E} can be attained only if $\theta\leq 120^\circ$.

To understand the case $\theta\geq 120^\circ$, I prove a slightly more general statement: If $X_1$, $X_2$ minimise~$E$, then they are the respective Fermat points~\cite{coxeter,johnson,gueron02} of triangles $AX_2D$, $BX_1C$. To show this, I assume to the contrary, and without loss of generality, that $X_1$ is different from the Fermat point $X_1'$ of triangle $AX_2D$. Moving $X_1\to X_1'$ gives a valid T1 configuration or ``Steiner tree'' because $X_1'$ is an interior point of $AX_2D$. This move reduces $E$ because, by definition of the Fermat point, ${|AX_1'|\!+\!|DX_1'|\!+\!|X_1'X_2|\!<\!|AX_1|\!+\!|DX_1|\!+\!|X_1X_2|}$, which contradicts minimality, so proves my claim. Now there are two possibilities: (1) ${X_1\not=X_2}$ and (2) $X_1=X_2=X$. In case (1), from properties of Fermat points~\cite{coxeter,johnson}, $X_1,X_2$ are the respective second intersections of the circumcircles of triangles $ADP_1$ and $BCP_2$ with lines $PX_2,PX_1$. This implies that $P_1,X_1,X_2,P_2$ lie on a line, whence, again, ${X_1=F_1}$, ${X_2=F_2}$, and hence $\theta\leq 120^\circ$ by the above. Case (2) must therefore arise for $\theta\geq120^\circ$. In that case, $\mathcal{L}=0$ and $\mathcal{E}_\text{b}=0$ by definition. The ``other'' T1 configuration is now the energy minimum; by taking $\theta\to180^\circ-\theta$, I note that it exists for $\theta\geq 60^\circ$. I will thus focus on $\theta\leq 120^\circ$ below.

By definition, the energy $\mathcal{E}$ at the T1 transition point~\figref{fig1}{c} minimises $|AX|+|BX|+|CX|+|DX|$ over all positions of the collapsed four-fold vertex $X$. It is well-known~\cite{saul} that this sum of the distances to the vertices of a convex quadrilateral is minimised at the intersection of its diagonals. This shows that $X=O$, i.e., $\mathcal{E}=|AO|+|BO|+|CO|+|DO|=|AC|+|BD|$. The energy barrier~\figref{fig1}{c} is the difference between this energy and the ground-state energy determined by Eq.~\eqref{eq:E} for $\theta\leq 120^\circ$. Hence $\mathcal{E}_\text{b}=|AC|+|BD|-|P_1P_2|$. Moreover, the distance from the T1 transition, i.e., the length of the inner edge, is ${\mathcal{L}=|F_1F_2|}$. I am left to express these quantities in terms of the geometry of quadrilateral $ABCD$~\figref{fig2}{a}. To compute $\mathcal{L}$, I rely, again, on properties of Fermat points~\cite{coxeter,johnson}: if I erect equilateral triangles $BF_1Q_1$ and $DF_2Q_2$ on segments $[BF_1]$ and $[DF_2]$, respectively, as shown in \textfigref{fig2}{d}, then lines $AQ_2$ and $CQ_1$ intersect line $P_1P_2$ at $F_1$ and $F_2$, respectively. It turns out to be convenient to take these computations to the complex plane. Using \textsc{Mathematica} (Wolfram, Inc.), I find~\footnote{See Supplemental Material at [url to be inserted], which includes Refs.~\cite{farhadifar07,staple10,fletcher14,fletcher16,alt17,grossman22,bi15,popovic21}, for details of the calculations and numerical results for a mean-field vertex model of an isolated T1 transition.}\nocite{farhadifar07,staple10,fletcher14,fletcher16,alt17,grossman22,bi15,popovic21}
\begin{widetext}
\vspace{-18pt}
\begin{subequations}\label{eq:EbL}
\begin{align}
\mathcal{E}_\text{b}&=\mathcal{D}_1+\mathcal{D}_2-\left[\mathcal{D}_1^2+\mathcal{D}_2^2+\bigl(\sqrt{3}\sin{\theta}-\cos{\theta}\bigr)\mathcal{D}_1\mathcal{D}_2\right]^{1/2},\\
\mathcal{L}&=\dfrac{\left(3\cos{\theta}+\sqrt{3}\sin{\theta}\right)^3}{9\left(2+\cos{2\theta}+\sqrt{3}\sin{2\theta}\right)}\dfrac{\mathcal{D}_1\mathcal{D}_2}{\left[\mathcal{D}_1^2+\mathcal{D}_2^2+\left(\sqrt{3}\sin{\theta}-\cos{\theta}\right)\mathcal{D}_1\mathcal{D}_2\right]^{1/2}}.
\end{align}
\end{subequations}
\vspace{-8pt}
\end{widetext}
Remarkably, these results depend on the geometry of the fixed outer vertices only through $\theta$, $\mathcal{D}_1=|AO|+|CO|$, $\mathcal{D}_2=|BO|+|DO|$, but are independent of $|AO|-|CO|$, $|BO|-|DO|$. Thus neither the energy barrier to nor the distance from the T1 transition are affected by changes, at constant $\theta$, of the relative position of the diagonals of the outer quadrilateral.

In the ensembles introduced in \textfigref{fig2}{b}, I can use $2\mathcal{A}=\mathcal{D}_1\mathcal{D}_2\sin{\theta}$ or $\mathcal{E}=\mathcal{D}_1+\mathcal{D}_2$ to express the dependence of Eqs.~\eqref{eq:EbL} on the geometry of this quadrilateral more symmetrically in terms of $\theta$ and $\Delta=|\mathcal{D}_1-\mathcal{D}_2|$ only. Since $\mathcal{D}_1=\mathcal{D}_2$ for a regular hexagonal lattice, $\Delta$ quantifies the geometric disorder of outer vertices.

Contour plots of $\mathcal{L}$ reveal that, in both the equal-area ensemble~\figref{fig2}{e} and the equal-energy ensemble~\figref{fig2}{f}, the distance from the T1 transition decreases as $\Delta$ increases at constant $\theta$. Similarly (not shown), the energy barrier decreases as $\Delta$ increases at constant $\theta$. Thus, in this tension-dominated regime, geometric disorder facilitates isolated T1 transitions. Both $\mathcal{E}_\text{b}$ and $\mathcal{L}$ vanish as $\theta\to120^\circ$ as expected [Figs.~\figrefp{fig2}{e} and \figrefp{fig2}{f}]. Expanding Eqs.~\eqref{eq:EbL} in this limit yields
\begin{align}
\dfrac{\mathcal{E}_\text{b}}{\mathcal{L}^2}\sim\dfrac{3}{8}\left(\dfrac{1}{\mathcal{D}_1}+\dfrac{1}{\mathcal{D}_2}\right)\quad\text{for }\theta\approx 120^\circ.
\end{align}
This displays the scaling $\mathcal{E}_\text{b}\sim\mathcal{L}^2$ that Ref.~\cite{popovic21} has shown to be generic. For larger $\mathcal{L}$, $\mathcal{E}_\text{b}$ increases faster than this scaling, in both ensembles [Figs.~\figrefp{fig2}{g} and~\figrefp{fig2}{h}], with the increase being larger at larger $\Delta$.

Next, I extend my results to the more general energy ${E=\gamma_a\ell_a+\gamma_b\ell_b+\gamma_c\ell_c+\gamma_d\ell_d+\gamma\ell}$, with (inhomogeneous) line tensions  $\gamma_a,\gamma_b,\gamma_c,\gamma_d,\gamma$ in the corresponding sides. This extends the above in the same way as the Fermat--Torricelli problem is extended by the Steiner problem~\cite{gueron02} of minimising the weighted sum of the distances to the vertices of a triangle, which is a special case of the Weber problem of operations research~\cite{church19,church22}. 

To minimise $E$ in this case, I erect a triangle $ADP_1$ such that ${|AD|:|P_1D|:|P_1A|=\gamma:\gamma_a:\gamma_d}$ outwards on line segment $[AD]$ and a triangle $BCP_2$ such that ${|BC|:|P_2C|:|P_2B|=\gamma:\gamma_b:\gamma_c}$ outwards on line segment $[BC]$. This assumes, of course, the existence of such triangles, and hence that the line tensions $\gamma_a,\gamma_b,\gamma_c,\gamma_d,\gamma$ satisfy the triangle inequalities, i.e., that they are not too different. This is probably a significant mathematical restriction of the geometric problem, but not perhaps a significant physical restriction of the biological problem, the more so as I will analyse small fluctuations around homogeneous tensions. I claim that the points $F_1$ and $F_2$ at which line $P_1P_2$ intersects the circumcircles of triangles $ADP_1$ and $BCP_2$ are again the positions of the inner vertices that minimise $E$. Indeed, Ptolemy's inequality~\cite{johnson}, applied to quadrilateral $AX_1DP_1$, yields ${\left|AX_1\right|\left|P_1D\right|+\left|DX_1\right|\left|P_1A\right|\geq\left|P_1X_1\right|\left|AD\right|}$, and hence, on scaling by $\left|AD\right|$, $\gamma_a|AX_1|+\gamma_d|DX_1|\geq \gamma|P_1X_1|$, with equality if and only if $A,X_1,D,P_1$ lie on a circle in this order. Similarly, $\gamma_b|BX_2|+\gamma_c|CX_2|\geq \gamma|P_2X_2|$, with equality if and only if $B,X_2,C,P_2$ are cocyclic. Thus
\begin{align}
E&=\bigl(\gamma_a|AX_1|+\gamma_d|DX_1|\bigr)+\gamma|X_1X_2|\nonumber\\
&\qquad+\bigl(\gamma_b|BX_2|+\gamma_c|CX_2|\bigr)\nonumber\\
&\geq \gamma\bigl(|P_1X_1|+|X_1X_2|+|P_2X_2|\bigr)\geq\gamma|P_1P_2|,
\end{align}
with equality, again, if and only if $X_1=F_1$ and $X_2=F_2$ and $P_1,F_1,F_2,P_2$ lie on a line in this order. By contrast to the homogeneous case, I am not aware of a simple condition equivalent to the last one. Again, this construction has additional geometric properties~\cite{[{The exact geometric solution of the Steiner problem~\cite{gueron02} involves a concurrency of lines and circles~\cite{johnson} that appears to have been discovered repeatedly, but has lately come to be known as Jacobi's theorem [see, e.g., }][{, }]kirby80,*[][{, }]gale96,*[][{]. This theorem shows that $F_1$ and $F_2$ are ``Steiner'' points of triangles $AF_2D$ and $BF_1C$, respectively, in the same way as they are Fermat points in the homogeneous case.}]vickers15}, but it does not lend itself to further general calculations.

It is therefore meet and right to specialise these results to the limit of small fluctuations, and analyse the energy barrier close to a T1 transition. Accordingly, I write $\theta=2\pi/3-\vartheta$, where $\vartheta\ll1$, and introduce fluctuations $\gamma_a=1+\vartheta\Gamma_a$, $\gamma_b=1+\vartheta\Gamma_b$, $\gamma_c=1+\vartheta\Gamma_c$, ${\gamma_d=1+\vartheta\Gamma_d}$, $\gamma=1+\vartheta\Gamma$, in which $\Gamma_a,\Gamma_b,\Gamma_c,\Gamma_d,\Gamma$ are independent, identically distributed fluctuations of zero mean and variance $\sigma^2$. For these fluctuations, I compute~\cite{Note1} the energy barrier $\mathcal{E}_\text{b}$ to a T1 transition by asymptotic expansion in~$\vartheta$. As I am not aware of an exact construction of the T1 transition point for this inhomogeneous case, I obtain~$\mathcal{E}$ by minimising energy order-by-order in $\vartheta$.

I thus find~\cite{Note1} that $\mathcal{E}_\text{b}=O\bigl(\vartheta^2\bigr)$, i.e., that the energy barrier from homogeneous tensions and the effect of fluctuations only arise at order $O\bigl(\vartheta^2\bigr)$. Continuing the asymptotic expansion to this order, I obtain~\cite{Note1}
\begin{widetext}
\vspace{-19pt}
\begin{subequations}
\begin{align}
\mathcal{E}_\text{b}&=\vartheta^2\left\{\dfrac{\mathcal{D}_1\mathcal{D}_2}{2(\mathcal{D}_1+\mathcal{D}_2)}+\dfrac{\left(\mathcal{D}_1\mathcal{D}_2\overline{\Gamma}+\mathcal{D}_1\delta_2\overline{\Gamma}_1+\mathcal{D}_2\delta_1\overline{\Gamma}_2\right)\left[\mathcal{D}_1\mathcal{D}_2\left(4\sqrt{3}+\overline{\Gamma}\right)+\mathcal{D}_1\delta_2\overline{\Gamma}_1+\mathcal{D}_2\delta_1\overline{\Gamma}_2\right]}{24\mathcal{D}_1\mathcal{D}_2(\mathcal{D}_1+\mathcal{D}_2)}\right\}+O\left(\vartheta^3\right),\label{eq:EbG}
\end{align}
with $\delta_1=|AO|\!-\!|CO|$, $\delta_2=|BO|\!-\!|DO|$, $\overline{\Gamma}=\Gamma_a\!+\!\Gamma_b\!+\!\Gamma_c\!+\!\Gamma_d\!-\!4\Gamma$, $\overline{\Gamma}_1=2\Gamma_c\!+\!\Gamma_d\!-\!2\Gamma_a\!-\!\Gamma_b$, ${\overline{\Gamma}_2=\Gamma_c\!+\!2\Gamma_d\!-\!\Gamma_a\!-\!2\Gamma_b}$.The first term in brackets is the leading-order energy barrier in the homogeneous case; the second term is associated with the tension fluctuations. It is clear that this second term can overcome the first, i.e., that small tension fluctuations can overcome the homogeneous energy barrier to drive T1 transitions. This is consistent with previous work highlighting the importance of tension fluctuations for tissue fluidisation~\cite{krajnc18,duclut21,curran17,tetley19,kim21,yamamoto22}. Averaging Eq.~\eqref{eq:EbG} yields
\begin{align}
\langle\mathcal{E}_\text{b}\rangle&=\vartheta^2\left[\dfrac{\mathcal{D}_1\mathcal{D}_2}{2(\mathcal{D}_1+\mathcal{D}_2)}+\dfrac{4(\mathcal{D}_1\delta_2\!+\!\mathcal{D}_2\delta_1)^2+\mathcal{D}_1^2\delta_2^2+\mathcal{D}_2^2\delta_1^2+10\mathcal{D}_1^2\mathcal{D}_2^2}{12\mathcal{D}_1\mathcal{D}_2(\mathcal{D}_1+\mathcal{D}_2)}\sigma^2\right]+O\left(\vartheta^3\right),
\end{align}
\end{subequations}
\vspace{-9pt}
\end{widetext}
showing that the \emph{average} correction from fluctuations is positive. Thus, on average, line-tension fluctuations increase the energy barrier to T1 transitions. This suggests in particular that slow fluctuations suppress T1 transitions compared to fast fluctuations because edges are in states in which their inhomogeneous tensions increase $\mathcal{E}_\text{b}$ most of the time. This is consistent with the numerical observations, in the vertex model~\cite{farhadifar07,staple10,fletcher14,fletcher16,alt17}, of Ref.~\cite{yamamoto22}.

\begin{figure}[b]
\includegraphics{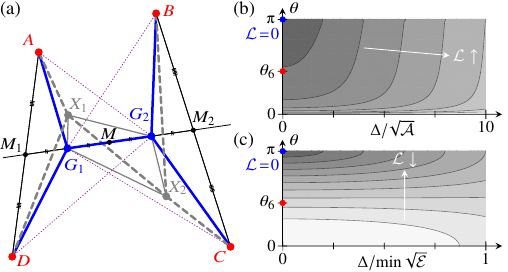}
\caption{Geometry of an isolated T1 transition dominated by (homogeneous) nonlinear line tensions. (a) Construction of the positions $G_1,G_2$ minimising the nonlinear line-tension energy for fixed outer vertices $A,B,C,D$. See text for further explanation. (b) Contour plot of the length $\mathcal{L}$ of the central edge against $\theta$ and the asymmetry $\Delta=|\mathcal{D}_2-\mathcal{D}_1|$ \figref{fig2}{a} in the equal-area ensemble [\textfigref{fig2}{b}, case (i)]. The red mark corresponds to a regular hexagonal lattice $\theta=\theta_6$, $\Delta=0$. The blue mark is the single point $\Delta=0$, $\theta=\pi$ at which ${\mathcal{L}=0}$. (c) Analogous contour plot in the equal-energy ensemble [\textfigref{fig2}{b}, case (ii)]. See text for further explanation.\label{fig3}\vspace{-5pt}}
\end{figure}

Finally, I turn to the case of (homogeneous) nonlinear line tensions, for which $E=\ell_a^2+\ell_b^2+\ell_c^2+\ell_d^2+\ell^2$. Nonlinear tensions of this ilk were introduced into tissue mechanics in Ref.~\cite{grossman22}. Unlike the standard perimeter elasticity in vertex models of tissues~\cite{farhadifar07,staple10,fletcher14,fletcher16,alt17,damavandi25}, they are local, in that $E$ can be expressed in terms of the lengths of the edges involved in the isolated T1 transition only. As shown in \textfigref{fig3}{a}, I denote by $M_1,M_2$ the respective midpoints of line segments $[AD],[BC]$, and divide line segment $[M_1M_2]$ into four equal segments by points $G_1,M,G_2$. By construction, $M_1G_2$ is a median of triangle $AG_2D$, and $|G_1G_2|=2|M_1G_1|$. It is well-known~\cite{coxeter,johnson} that this implies that $G_1$ is the centroid of triangle $AG_2D$. Analogously, $G_2$ is the centroid of triangle $BG_1C$. I claim that $G_1, G_2$ are now the positions of the inner vertices that minimise $E$. Thus, in this nonlinear case, the pair of Fermat points of the linear case is replaced by a pair of centroids. To prove this, I shall need an arcane piece of triangle trivia~\cite{johnson}:
\begin{lemma}
Let $ABC$ be a triangle with centroid $G$. Then $|PA|^2+|PB|^2+|PC|^2=|GA|^2+|GB|^2+|GC|^2+3|GP|^2$ for any point $P$.
\end{lemma}
\noindent Let $X_1,X_2$ be points in the plane. Applying this result to points $X_1,X_2$ in triangles $AG_2D,BG_1C$, respectively,
\begin{align}
E&=\bigl(|AX_1|^2\!+\!|DX_1|^2\bigr)+\bigl(|BX_2|^2\!+\!|CX_2|^2\bigr)+|X_1X_2|^2\nonumber\\
&=\bigl(|AG_1|^2\!+\!|DG_1|^2\!+\!|G_1G_2|^2\!+\!3|G_1X_1|^2\!-\!|X_1G_2|^2\bigr)\nonumber\\
&\quad+\bigl(|BG_2|^2\!+\!|CG_2|^2\!+\!|G_1G_2|^2\!+\!3|G_2X_2|^2\!-\!|X_2G_1|^2\bigr)\nonumber\\
&\qquad +|X_1X_2|^2\nonumber\\
&=|AG_1|^2+|DG_1|^2+|BG_2|^2+|CG_2|^2+|G_1G_2|^2\nonumber\\
&\quad+\bigl(|X_1X_2|^2+|X_2G_2|^2+|G_1G_2|^2+|G_1X_1|^2\nonumber\\
&\hspace{10mm}-|X_1G_2|^2-|X_2G_1|^2\bigr)+2\bigl(|G_1X_1|^2+|G_2X_2|^2\bigr)\nonumber\\
&\geq |AG_1|^2\!+\!|DG_1|^2+|BG_2|^2\!+\!|CG_2|^2+|G_1G_2|^2,\hspace{-5pt}
\end{align}
in which I have applied Euler's quadrilateral theorem \mbox{\cite{johnson,dunham}} to quadrilateral $X_1X_2G_2G_1$ to note that the terms in parentheses in the penultimate equation are nonnegative. This completes the proof.

It is now clear that $\mathcal{L}=|G_1G_2|=|M_1M_2|/2$, which cannot vanish if $ABCD$ is convex. This establishes that nonlinear tensions resist T1 transitions. Next, I introduce the barycentre $Z$ of quadrilateral $ABCD$. It minimises the sum of the squares of the distances to its vertices~\footnote{This fact is surely folklore, but I could not find an exact reference. To prove it, I denote by $\vec{r_A},\vec{r_B},\vec{r_C},\vec{r_D}$ the position vectors of 
$A,B,C,D$ and compute the gradient of $E(\vec{r})=\|\vec{r}-\vec{r_A}\|^2+\|\vec{r}-\vec{r_B}\|^2+\|\vec{r}-\vec{r_C}\|^2+\|\vec{r}-\vec{r_D}\|^2$, $\vec{\nabla}E=2\bigl(4\vec{r}-\vec{r_A}-\vec{r_B}-\vec{r_C}-\vec{r_D}\bigr)$, which has a single stationary point at $\vec{r}=(\vec{r_A}+\vec{r_B}+\vec{r_C}+\vec{r_D})/4$, i.e., at the barycentre, and this is clearly a minimum.}, so that the energy at the T1 transition point~\figref{fig1}{c} is ${\mathcal{E}=|AZ|^2+|BZ|^2+|CZ|^2+|DZ|^2}$. With this result, I can compute~\cite{Note1}
\begin{align}
\mathcal{E}_\text{b}=\dfrac{\mathcal{D}_1^2+\mathcal{D}_2^2+2\mathcal{D}_1\mathcal{D}_2\cos{\theta}}{8}=\dfrac{\mathcal{L}^2}{2},
\end{align}
so $\mathcal{E}_\text{b}=\mathcal{L}=0$ in and only in the degenerate case $\theta=\pi$, $\mathcal{D}_1=\mathcal{D}_2$. Again, these results only depend on the diagonals $\mathcal{D}_1,\mathcal{D}_2$ of the quadrilateral formed by the outer vertices of the T1 transition and the angle $\theta$ between them \figref{fig2}{a}. Interestingly, the generic scaling $\mathcal{E}_\text{b}\sim\mathcal{L}^2$ for small $\mathcal{L}$~\cite{popovic21} is an exact proportionality in this case. In the equal-area ensemble \mbox{[\textfigref{fig2}{b}, case (i)]}, $\mathcal{L}$ increases with the disorder $\Delta=|\mathcal{D}_1-\mathcal{D}_2|$ of the outer quadrilateral \figref{fig3}{b}. The equal-energy ensemble [\textfigref{fig2}{b}, case (ii)] is more interesting, because $\mathcal{E}$ also depends on $\delta_1=|AO|\!-\!|CO|$, $\delta_2=|BO|\!-\!|DO|$, viz., 
\begin{align}
\mathcal{E}=\dfrac{\mathcal{D}_1^2+\mathcal{D}_2^2}{2}+\dfrac{\delta_1^2+\delta_2^2+2\delta_1\delta_2\cos{\theta}}{4}.
\end{align}
I therefore rescale $\Delta$ with $\min{\mathcal{E}}=\bigl(\mathcal{D}_1^2+\mathcal{D}_2^2\bigr)/2$, attained for $\delta_1=\delta_2=0$, to plot the contours of $\mathcal{L}$ in \textfigref{fig3}{c}. Now $\mathcal{L}$ increases with $\Delta$ only for large enough~$\theta$, for which $\mathcal{L}$ is small and hence the energy barrier to a T1 transition is lower. Hence disorder further resists T1 transitions in this nonlinear case.

I have thus shown, by relying on exact geometric constructions, how geometry, fluctuations, and nonlinearities control general isolated, line-tension-dominated T1 transitions in epithelial tissues. Even these simple models have revealed complex behaviour.

The linear and nonlinear line tensions that I have studied here are different, however, from the perimeter elasticity in the single-cell energy of vertex models of tissues~\cite{farhadifar07,staple10,fletcher14,fletcher16,alt17,damavandi25}. Moreover, these vertex models feature additional interplay between this perimeter elasticity and area elasticity. In the Supplemental Material~\cite{Note1}, I therefore analyse a mean-field vertex model of an isolated T1 transition. My results confirm that the distance $\mathcal{L}$ from a T1 transition increases with increasing disorder in general, but also uncover more complex behaviour in some parameter regimes~\cite{Note1}. Future work will need to explain these numerical observations, in particular by generalisations of my geometric constructions to cases in which area elasticity dominates.

In general, the energy barrier $\mathcal{E}_\text{b}=\mathcal{E}_\text{b}^\text{local}-\mathcal{E}^\text{nonlocal}_\text{b}$ decomposes into a local contribution that is the energy barrier for an isolated T1 transition and a nonlocal correction resulting from the displacement of the vertices neighbouring those undergoing the T1 transition as the surrounding cells deform. Active T1 transitions \cite{krajnc18,krajnc20,krajnc21,duclut22,sknepnek23,brauns24} overcome this local energy barrier. Nonlocal relaxation may drive additional, passive rearrangements that cannot be captured by this local picture. In particular, the mean-field arguments that predict fluidisation in the vertex model even quantitatively \cite{bi15,sahu20,staddon25} do rely on such nonlocal effects. Nonetheless, recent numerical simulations of the vertex model~\cite{demarzio25} have suggested that these two contributions are proportional to each other, ${\mathcal{E}^\text{nonlocal}_\text{b}\propto\mathcal{E}_\text{b}^\text{local}}$. While the mechanism setting this proportionality remains unclear, this suggests that mean-field models of isolated T1 transitions that I have studied here capture the full physics of T1 transitions in tissues and that they can be extended into quantitative models of T1 transitions by including this proportionality. 

In this context, a large body of recent work has focused on the continuum mechanics of epithelial tissues, computing the viscoelastic moduli of the vertex model~\cite{murisic15,nestorbergmann17,tong22,hernandez22,tong23,staddon23,hernandez23,kim24} and deriving coarse-grained limits of the vertex model~\cite{grossman22,fielding23,perez23,triguero23,grossman25} and of differential-tension models of tissues~\cite{krajnc15,haas19,andrensek23,andrensek25}. My geometric results form the basis for understanding the mean-field effects of cell-scale disorder on cell rearrangements in these models and in related continuum theories~\cite{claussen24} for tissues. At the same time, three-dimensional vertex models of bulk tissues~\cite{honda04,sahu21,sanematsu21,zhang22,villeneuve24} and descriptions of their cell rearrangements~\cite{sarkar24} are beginning to emerge, but understanding their coarse-grained mechanics remains a formidable challenge for future work.
\begin{acknowledgments}
I thank K. Chhajed, M. Popovi\'c, and M. F. Staddon for discussions or comments on the manuscript. I am grateful for funding from the Max Planck Society.
\end{acknowledgments}

\bibliography{main}

\begin{thebibliography}{9}%
\makeatletter
\providecommand \@ifxundefined [1]{%
 \@ifx{#1\undefined}
}%
\providecommand \@ifnum [1]{%
 \ifnum #1\expandafter \@firstoftwo
 \else \expandafter \@secondoftwo
 \fi
}%
\providecommand \@ifx [1]{%
 \ifx #1\expandafter \@firstoftwo
 \else \expandafter \@secondoftwo
 \fi
}%
\providecommand \natexlab [1]{#1}%
\providecommand \enquote  [1]{``#1''}%
\providecommand \bibnamefont  [1]{#1}%
\providecommand \bibfnamefont [1]{#1}%
\providecommand \citenamefont [1]{#1}%
\providecommand \href@noop [0]{\@secondoftwo}%
\providecommand \href [0]{\begingroup \@sanitize@url \@href}%
\providecommand \@href[1]{\@@startlink{#1}\@@href}%
\providecommand \@@href[1]{\endgroup#1\@@endlink}%
\providecommand \@sanitize@url [0]{\catcode `\\12\catcode `\$12\catcode
  `\&12\catcode `\#12\catcode `\^12\catcode `\_12\catcode `\%12\relax}%
\providecommand \@@startlink[1]{}%
\providecommand \@@endlink[0]{}%
\providecommand \url  [0]{\begingroup\@sanitize@url \@url }%
\providecommand \@url [1]{\endgroup\@href {#1}{\urlprefix }}%
\providecommand \urlprefix  [0]{URL }%
\providecommand \Eprint [0]{\href }%
\providecommand \doibase [0]{https://doi.org/}%
\providecommand \selectlanguage [0]{\@gobble}%
\providecommand \bibinfo  [0]{\@secondoftwo}%
\providecommand \bibfield  [0]{\@secondoftwo}%
\providecommand \translation [1]{[#1]}%
\providecommand \BibitemOpen [0]{}%
\providecommand \bibitemStop [0]{}%
\providecommand \bibitemNoStop [0]{.\EOS\space}%
\providecommand \EOS [0]{\spacefactor3000\relax}%
\providecommand \BibitemShut  [1]{\csname bibitem#1\endcsname}%
\let\auto@bib@innerbib\@empty
\bibitem [{\citenamefont {Farhadifar}\ \emph {et~al.}(2007)\citenamefont
  {Farhadifar}, \citenamefont {Röper}, \citenamefont {Aigouy}, \citenamefont
  {Eaton},\ and\ \citenamefont {Jülicher}}]{farhadifar07}%
  \BibitemOpen
  \bibfield  {author} {\bibinfo {author} {\bibfnamefont {R.}~\bibnamefont
  {Farhadifar}}, \bibinfo {author} {\bibfnamefont {J.-C.}\ \bibnamefont
  {Röper}}, \bibinfo {author} {\bibfnamefont {B.}~\bibnamefont {Aigouy}},
  \bibinfo {author} {\bibfnamefont {S.}~\bibnamefont {Eaton}},\ and\ \bibinfo
  {author} {\bibfnamefont {F.}~\bibnamefont {Jülicher}},\ }\bibfield  {title}
  {\bibinfo {title} {The influence of cell mechanics, cell-cell interactions,
  and proliferation on epithelial packing},\ }\href
  {https://doi.org/10.1016/j.cub.2007.11.049} {\bibfield  {journal} {\bibinfo
  {journal} {Curr. Biol.}\ }\textbf {\bibinfo {volume} {17}},\ \bibinfo {pages}
  {2095} (\bibinfo {year} {2007})}\BibitemShut {NoStop}%
\bibitem [{\citenamefont {Staple}\ \emph {et~al.}(2010)\citenamefont {Staple},
  \citenamefont {Farhadifar}, \citenamefont {R{\"o}per}, \citenamefont
  {Aigouy}, \citenamefont {Eaton},\ and\ \citenamefont
  {J{\"u}licher}}]{staple10}%
  \BibitemOpen
  \bibfield  {author} {\bibinfo {author} {\bibfnamefont {D.~B.}\ \bibnamefont
  {Staple}}, \bibinfo {author} {\bibfnamefont {R.}~\bibnamefont {Farhadifar}},
  \bibinfo {author} {\bibfnamefont {J.-C.}\ \bibnamefont {R{\"o}per}}, \bibinfo
  {author} {\bibfnamefont {B.}~\bibnamefont {Aigouy}}, \bibinfo {author}
  {\bibfnamefont {S.}~\bibnamefont {Eaton}},\ and\ \bibinfo {author}
  {\bibfnamefont {F.}~\bibnamefont {J{\"u}licher}},\ }\bibfield  {title}
  {\bibinfo {title} {Mechanics and remodelling of cell packings in epithelia},\
  }\href {https://doi.org/10.1140/epje/i2010-10677-0} {\bibfield  {journal}
  {\bibinfo  {journal} {Eur. Phys. J. E}\ }\textbf {\bibinfo {volume} {33}},\
  \bibinfo {pages} {117} (\bibinfo {year} {2010})}\BibitemShut {NoStop}%
\bibitem [{\citenamefont {Fletcher}\ \emph {et~al.}(2014)\citenamefont
  {Fletcher}, \citenamefont {Osterfield}, \citenamefont {Baker},\ and\
  \citenamefont {Shvartsman}}]{fletcher14}%
  \BibitemOpen
  \bibfield  {author} {\bibinfo {author} {\bibfnamefont {A.~G.}\ \bibnamefont
  {Fletcher}}, \bibinfo {author} {\bibfnamefont {M.}~\bibnamefont
  {Osterfield}}, \bibinfo {author} {\bibfnamefont {R.~E.}\ \bibnamefont
  {Baker}},\ and\ \bibinfo {author} {\bibfnamefont {S.~Y.}\ \bibnamefont
  {Shvartsman}},\ }\bibfield  {title} {\bibinfo {title} {Vertex models of
  epithelial morphogenesis},\ }\href
  {https://doi.org/10.1016/j.bpj.2013.11.4498} {\bibfield  {journal} {\bibinfo
  {journal} {Biophys. J.}\ }\textbf {\bibinfo {volume} {106}},\ \bibinfo
  {pages} {2291} (\bibinfo {year} {2014})}\BibitemShut {NoStop}%
\bibitem [{\citenamefont {Fletcher}\ \emph {et~al.}(2016)\citenamefont
  {Fletcher}, \citenamefont {Cooper},\ and\ \citenamefont
  {Baker}}]{fletcher16}%
  \BibitemOpen
  \bibfield  {author} {\bibinfo {author} {\bibfnamefont {A.~G.}\ \bibnamefont
  {Fletcher}}, \bibinfo {author} {\bibfnamefont {F.}~\bibnamefont {Cooper}},\
  and\ \bibinfo {author} {\bibfnamefont {R.~E.}\ \bibnamefont {Baker}},\
  }\bibfield  {title} {\bibinfo {title} {Mechanocellular models of epithelial
  morphogenesis},\ }\href {https://doi.org/10.1098/rstb.2015.0519} {\bibfield
  {journal} {\bibinfo  {journal} {Phil. Trans. Roy. Soc. B}\ }\textbf {\bibinfo
  {volume} {372}},\ \bibinfo {pages} {20150519} (\bibinfo {year}
  {2016})}\BibitemShut {NoStop}%
\bibitem [{\citenamefont {Alt}\ \emph {et~al.}(2017)\citenamefont {Alt},
  \citenamefont {Ganguly},\ and\ \citenamefont {Salbreux}}]{alt17}%
  \BibitemOpen
  \bibfield  {author} {\bibinfo {author} {\bibfnamefont {S.}~\bibnamefont
  {Alt}}, \bibinfo {author} {\bibfnamefont {P.}~\bibnamefont {Ganguly}},\ and\
  \bibinfo {author} {\bibfnamefont {G.}~\bibnamefont {Salbreux}},\ }\bibfield
  {title} {\bibinfo {title} {Vertex models: from cell mechanics to tissue
  morphogenesis},\ }\href {https://doi.org/10.1098/rstb.2015.0520} {\bibfield
  {journal} {\bibinfo  {journal} {Phil. Trans. Roy. Soc. B}\ }\textbf {\bibinfo
  {volume} {372}},\ \bibinfo {pages} {20150520} (\bibinfo {year}
  {2017})}\BibitemShut {NoStop}%
\bibitem [{\citenamefont {Grossman}\ and\ \citenamefont
  {Joanny}(2022)}]{grossman22}%
  \BibitemOpen
  \bibfield  {author} {\bibinfo {author} {\bibfnamefont {D.}~\bibnamefont
  {Grossman}}\ and\ \bibinfo {author} {\bibfnamefont {J.-F.}\ \bibnamefont
  {Joanny}},\ }\bibfield  {title} {\bibinfo {title} {Instabilities and geometry
  of growing tissues},\ }\href {https://doi.org/10.1103/PhysRevLett.129.048102}
  {\bibfield  {journal} {\bibinfo  {journal} {Phys. Rev. Lett.}\ }\textbf
  {\bibinfo {volume} {129}},\ \bibinfo {pages} {048102} (\bibinfo {year}
  {2022})}\BibitemShut {NoStop}%
\bibitem [{Note1()}]{Note1}%
  \BibitemOpen
  \bibinfo {note} {\protect \textsc {Matlab} code for minimization of this
  mean-field vertex model energy is available online at \protect \texttt
  {zenodo.org/...}.}\BibitemShut {Stop}%
\bibitem [{\citenamefont {Bi}\ \emph {et~al.}(2015)\citenamefont {Bi},
  \citenamefont {Lopez}, \citenamefont {Schwarz},\ and\ \citenamefont
  {Manning}}]{bi15}%
  \BibitemOpen
  \bibfield  {author} {\bibinfo {author} {\bibfnamefont {D.}~\bibnamefont
  {Bi}}, \bibinfo {author} {\bibfnamefont {J.~H.}\ \bibnamefont {Lopez}},
  \bibinfo {author} {\bibfnamefont {J.~M.}\ \bibnamefont {Schwarz}},\ and\
  \bibinfo {author} {\bibfnamefont {M.~L.}\ \bibnamefont {Manning}},\
  }\bibfield  {title} {\bibinfo {title} {A density-independent rigidity
  transition in biological tissues},\ }\href
  {https://doi.org/10.1038/nphys3471} {\bibfield  {journal} {\bibinfo
  {journal} {Nat. Phys.}\ }\textbf {\bibinfo {volume} {11}},\ \bibinfo {pages}
  {1074} (\bibinfo {year} {2015})}\BibitemShut {NoStop}%
\bibitem [{\citenamefont {Popović}\ \emph {et~al.}(2021)\citenamefont
  {Popović}, \citenamefont {Druelle}, \citenamefont {Dye}, \citenamefont
  {Jülicher},\ and\ \citenamefont {Wyart}}]{popovic21}%
  \BibitemOpen
  \bibfield  {author} {\bibinfo {author} {\bibfnamefont {M.}~\bibnamefont
  {Popović}}, \bibinfo {author} {\bibfnamefont {V.}~\bibnamefont {Druelle}},
  \bibinfo {author} {\bibfnamefont {N.~A.}\ \bibnamefont {Dye}}, \bibinfo
  {author} {\bibfnamefont {F.}~\bibnamefont {Jülicher}},\ and\ \bibinfo
  {author} {\bibfnamefont {M.}~\bibnamefont {Wyart}},\ }\bibfield  {title}
  {\bibinfo {title} {Inferring the flow properties of epithelial tissues from
  their geometry},\ }\href {https://doi.org/10.1088/1367-2630/abcbc7}
  {\bibfield  {journal} {\bibinfo  {journal} {New J. Phys.}\ }\textbf {\bibinfo
  {volume} {23}},\ \bibinfo {pages} {033004} (\bibinfo {year}
  {2021})}\BibitemShut {NoStop}%
\end{thebibliography}%


\begin{thebibliography}{84}%
\makeatletter
\providecommand \@ifxundefined [1]{%
 \@ifx{#1\undefined}
}%
\providecommand \@ifnum [1]{%
 \ifnum #1\expandafter \@firstoftwo
 \else \expandafter \@secondoftwo
 \fi
}%
\providecommand \@ifx [1]{%
 \ifx #1\expandafter \@firstoftwo
 \else \expandafter \@secondoftwo
 \fi
}%
\providecommand \natexlab [1]{#1}%
\providecommand \enquote  [1]{``#1''}%
\providecommand \bibnamefont  [1]{#1}%
\providecommand \bibfnamefont [1]{#1}%
\providecommand \citenamefont [1]{#1}%
\providecommand \href@noop [0]{\@secondoftwo}%
\providecommand \href [0]{\begingroup \@sanitize@url \@href}%
\providecommand \@href[1]{\@@startlink{#1}\@@href}%
\providecommand \@@href[1]{\endgroup#1\@@endlink}%
\providecommand \@sanitize@url [0]{\catcode `\\12\catcode `\$12\catcode
  `\&12\catcode `\#12\catcode `\^12\catcode `\_12\catcode `\%12\relax}%
\providecommand \@@startlink[1]{}%
\providecommand \@@endlink[0]{}%
\providecommand \url  [0]{\begingroup\@sanitize@url \@url }%
\providecommand \@url [1]{\endgroup\@href {#1}{\urlprefix }}%
\providecommand \urlprefix  [0]{URL }%
\providecommand \Eprint [0]{\href }%
\providecommand \doibase [0]{https://doi.org/}%
\providecommand \selectlanguage [0]{\@gobble}%
\providecommand \bibinfo  [0]{\@secondoftwo}%
\providecommand \bibfield  [0]{\@secondoftwo}%
\providecommand \translation [1]{[#1]}%
\providecommand \BibitemOpen [0]{}%
\providecommand \bibitemStop [0]{}%
\providecommand \bibitemNoStop [0]{.\EOS\space}%
\providecommand \EOS [0]{\spacefactor3000\relax}%
\providecommand \BibitemShut  [1]{\csname bibitem#1\endcsname}%
\let\auto@bib@innerbib\@empty
\bibitem [{\citenamefont {Walck-Shannon}\ and\ \citenamefont
  {Hardin}(2014)}]{walck14}%
  \BibitemOpen
  \bibfield  {author} {\bibinfo {author} {\bibfnamefont {E.}~\bibnamefont
  {Walck-Shannon}}\ and\ \bibinfo {author} {\bibfnamefont {J.}~\bibnamefont
  {Hardin}},\ }\bibfield  {title} {\bibinfo {title} {Cell intercalation from
  top to bottom},\ }\href {https://doi.org/10.1038/nrm3723} {\bibfield
  {journal} {\bibinfo  {journal} {Nat. Rev. Mol. Cell Biol.}\ }\textbf
  {\bibinfo {volume} {15}},\ \bibinfo {pages} {34} (\bibinfo {year}
  {2014})}\BibitemShut {NoStop}%
\bibitem [{\citenamefont {Rauzi}(2020)}]{rauzi20}%
  \BibitemOpen
  \bibfield  {author} {\bibinfo {author} {\bibfnamefont {M.}~\bibnamefont
  {Rauzi}},\ }\bibfield  {title} {\bibinfo {title} {Cell intercalation in a
  simple epithelium},\ }\href {https://doi.org/10.1098/rstb.2019.0552}
  {\bibfield  {journal} {\bibinfo  {journal} {Phil. Trans. Roy. Soc. B}\
  }\textbf {\bibinfo {volume} {375}},\ \bibinfo {pages} {20190552} (\bibinfo
  {year} {2020})}\BibitemShut {NoStop}%
\bibitem [{\citenamefont {Wolpert}\ \emph {et~al.}(2000)\citenamefont
  {Wolpert}, \citenamefont {Smith}, \citenamefont {Keller}, \citenamefont
  {Davidson}, \citenamefont {Edlund}, \citenamefont {Elul}, \citenamefont
  {Ezin}, \citenamefont {Shook},\ and\ \citenamefont {Skoglund}}]{wolpert00}%
  \BibitemOpen
  \bibfield  {author} {\bibinfo {author} {\bibfnamefont {L.}~\bibnamefont
  {Wolpert}}, \bibinfo {author} {\bibfnamefont {J.~C.}\ \bibnamefont {Smith}},
  \bibinfo {author} {\bibfnamefont {R.}~\bibnamefont {Keller}}, \bibinfo
  {author} {\bibfnamefont {L.}~\bibnamefont {Davidson}}, \bibinfo {author}
  {\bibfnamefont {A.}~\bibnamefont {Edlund}}, \bibinfo {author} {\bibfnamefont
  {T.}~\bibnamefont {Elul}}, \bibinfo {author} {\bibfnamefont {M.}~\bibnamefont
  {Ezin}}, \bibinfo {author} {\bibfnamefont {D.}~\bibnamefont {Shook}},\ and\
  \bibinfo {author} {\bibfnamefont {P.}~\bibnamefont {Skoglund}},\ }\bibfield
  {title} {\bibinfo {title} {Mechanisms of convergence and extension by cell
  intercalation},\ }\href {https://doi.org/10.1098/rstb.2000.0626} {\bibfield
  {journal} {\bibinfo  {journal} {Phil. Trans. Roy. Soc. B}\ }\textbf {\bibinfo
  {volume} {355}},\ \bibinfo {pages} {897} (\bibinfo {year}
  {2000})}\BibitemShut {NoStop}%
\bibitem [{\citenamefont {Wallingford}\ \emph {et~al.}(2002)\citenamefont
  {Wallingford}, \citenamefont {Fraser},\ and\ \citenamefont
  {Harland}}]{wallingford02}%
  \BibitemOpen
  \bibfield  {author} {\bibinfo {author} {\bibfnamefont {J.~B.}\ \bibnamefont
  {Wallingford}}, \bibinfo {author} {\bibfnamefont {S.~E.}\ \bibnamefont
  {Fraser}},\ and\ \bibinfo {author} {\bibfnamefont {R.~M.}\ \bibnamefont
  {Harland}},\ }\bibfield  {title} {\bibinfo {title} {Convergent extension: The
  molecular control of polarized cell movement during embryonic development},\
  }\href {https://doi.org/10.1016/S1534-5807(02)00197-1} {\bibfield  {journal}
  {\bibinfo  {journal} {Dev. Cell}\ }\textbf {\bibinfo {volume} {2}},\ \bibinfo
  {pages} {695} (\bibinfo {year} {2002})}\BibitemShut {NoStop}%
\bibitem [{\citenamefont {Irvine}\ and\ \citenamefont
  {Wieschaus}(1994)}]{irvine94}%
  \BibitemOpen
  \bibfield  {author} {\bibinfo {author} {\bibfnamefont {K.~D.}\ \bibnamefont
  {Irvine}}\ and\ \bibinfo {author} {\bibfnamefont {E.}~\bibnamefont
  {Wieschaus}},\ }\bibfield  {title} {\bibinfo {title} {Cell intercalation
  during \emph{Drosophila} germband extension and its regulation by pair-rule
  segmentation genes},\ }\href {https://doi.org/10.1242/dev.120.4.827}
  {\bibfield  {journal} {\bibinfo  {journal} {Development}\ }\textbf {\bibinfo
  {volume} {120}},\ \bibinfo {pages} {827} (\bibinfo {year}
  {1994})}\BibitemShut {NoStop}%
\bibitem [{\citenamefont {Bertet}\ \emph {et~al.}(2004)\citenamefont {Bertet},
  \citenamefont {Sulak},\ and\ \citenamefont {Lecuit}}]{bertet04}%
  \BibitemOpen
  \bibfield  {author} {\bibinfo {author} {\bibfnamefont {C.}~\bibnamefont
  {Bertet}}, \bibinfo {author} {\bibfnamefont {L.}~\bibnamefont {Sulak}},\ and\
  \bibinfo {author} {\bibfnamefont {T.}~\bibnamefont {Lecuit}},\ }\bibfield
  {title} {\bibinfo {title} {Myosin-dependent junction remodelling controls
  planar cell intercalation and axis elongation},\ }\href
  {https://doi.org/10.1038/nature02590} {\bibfield  {journal} {\bibinfo
  {journal} {Nature}\ }\textbf {\bibinfo {volume} {429}},\ \bibinfo {pages}
  {667} (\bibinfo {year} {2004})}\BibitemShut {NoStop}%
\bibitem [{\citenamefont {Kong}\ \emph {et~al.}(2017)\citenamefont {Kong},
  \citenamefont {Wolf},\ and\ \citenamefont {Großhans}}]{kong17}%
  \BibitemOpen
  \bibfield  {author} {\bibinfo {author} {\bibfnamefont {D.}~\bibnamefont
  {Kong}}, \bibinfo {author} {\bibfnamefont {F.}~\bibnamefont {Wolf}},\ and\
  \bibinfo {author} {\bibfnamefont {J.}~\bibnamefont {Großhans}},\ }\bibfield
  {title} {\bibinfo {title} {Forces directing germ-band extension in
  \emph{Drosophila} embryos},\ }\href
  {https://doi.org/10.1016/j.mod.2016.12.001} {\bibfield  {journal} {\bibinfo
  {journal} {Mech. Dev.}\ }\textbf {\bibinfo {volume} {144}},\ \bibinfo {pages}
  {11} (\bibinfo {year} {2017})}\BibitemShut {NoStop}%
\bibitem [{\citenamefont {Brauns}\ \emph {et~al.}(2024)\citenamefont {Brauns},
  \citenamefont {Claussen}, \citenamefont {Lefebvre}, \citenamefont
  {Wieschaus},\ and\ \citenamefont {Shraiman}}]{brauns24}%
  \BibitemOpen
  \bibfield  {author} {\bibinfo {author} {\bibfnamefont {F.}~\bibnamefont
  {Brauns}}, \bibinfo {author} {\bibfnamefont {N.~H.}\ \bibnamefont
  {Claussen}}, \bibinfo {author} {\bibfnamefont {M.~F.}\ \bibnamefont
  {Lefebvre}}, \bibinfo {author} {\bibfnamefont {E.~F.}\ \bibnamefont
  {Wieschaus}},\ and\ \bibinfo {author} {\bibfnamefont {B.~I.}\ \bibnamefont
  {Shraiman}},\ }\bibfield  {title} {\bibinfo {title} {The geometric basis of
  epithelial convergent extension},\ }\href
  {https://doi.org/10.7554/eLife.95521} {\bibfield  {journal} {\bibinfo
  {journal} {eLife}\ }\textbf {\bibinfo {volume} {13}},\ \bibinfo {pages}
  {RP95521} (\bibinfo {year} {2024})}\BibitemShut {NoStop}%
\bibitem [{\citenamefont {Cantat}\ \emph {et~al.}(2013)\citenamefont {Cantat},
  \citenamefont {Cohen-Addad}, \citenamefont {Elias}, \citenamefont {Graner},
  \citenamefont {Höhler}, \citenamefont {Pitois}, \citenamefont {Rouyer},\
  and\ \citenamefont {Saint-Jalmes}}]{cantat}%
  \BibitemOpen
  \bibfield  {author} {\bibinfo {author} {\bibfnamefont {I.}~\bibnamefont
  {Cantat}}, \bibinfo {author} {\bibfnamefont {S.}~\bibnamefont {Cohen-Addad}},
  \bibinfo {author} {\bibfnamefont {F.}~\bibnamefont {Elias}}, \bibinfo
  {author} {\bibfnamefont {F.}~\bibnamefont {Graner}}, \bibinfo {author}
  {\bibfnamefont {R.}~\bibnamefont {Höhler}}, \bibinfo {author} {\bibfnamefont
  {O.}~\bibnamefont {Pitois}}, \bibinfo {author} {\bibfnamefont
  {F.}~\bibnamefont {Rouyer}},\ and\ \bibinfo {author} {\bibfnamefont
  {A.}~\bibnamefont {Saint-Jalmes}},\ }\href
  {https://doi.org/10.1093/acprof:oso/9780199662890.003.0003} {\emph {\bibinfo
  {title} {Foams: Structure and Dynamics}}}\ (\bibinfo  {publisher} {Oxford
  University Press},\ \bibinfo {address} {Oxford, UK},\ \bibinfo {year}
  {2013})\ Chap.~\bibinfo {chapter} {3}, pp.\ \bibinfo {pages}
  {75--131}\BibitemShut {NoStop}%
\bibitem [{\citenamefont {Bi}\ \emph {et~al.}(2014)\citenamefont {Bi},
  \citenamefont {Lopez}, \citenamefont {Schwarz},\ and\ \citenamefont
  {Manning}}]{bi14}%
  \BibitemOpen
  \bibfield  {author} {\bibinfo {author} {\bibfnamefont {D.}~\bibnamefont
  {Bi}}, \bibinfo {author} {\bibfnamefont {J.~H.}\ \bibnamefont {Lopez}},
  \bibinfo {author} {\bibfnamefont {J.~M.}\ \bibnamefont {Schwarz}},\ and\
  \bibinfo {author} {\bibfnamefont {M.~L.}\ \bibnamefont {Manning}},\
  }\bibfield  {title} {\bibinfo {title} {Energy barriers and cell migration in
  densely packed tissues},\ }\href {https://doi.org/10.1039/C3SM52893F}
  {\bibfield  {journal} {\bibinfo  {journal} {Soft Matter}\ }\textbf {\bibinfo
  {volume} {10}},\ \bibinfo {pages} {1885} (\bibinfo {year}
  {2014})}\BibitemShut {NoStop}%
\bibitem [{\citenamefont {Krajnc}\ \emph {et~al.}(2018)\citenamefont {Krajnc},
  \citenamefont {Dasgupta}, \citenamefont {Ziherl},\ and\ \citenamefont
  {Prost}}]{krajnc18}%
  \BibitemOpen
  \bibfield  {author} {\bibinfo {author} {\bibfnamefont {M.}~\bibnamefont
  {Krajnc}}, \bibinfo {author} {\bibfnamefont {S.}~\bibnamefont {Dasgupta}},
  \bibinfo {author} {\bibfnamefont {P.}~\bibnamefont {Ziherl}},\ and\ \bibinfo
  {author} {\bibfnamefont {J.}~\bibnamefont {Prost}},\ }\bibfield  {title}
  {\bibinfo {title} {Fluidization of epithelial sheets by active cell
  rearrangements},\ }\href {https://doi.org/10.1103/PhysRevE.98.022409}
  {\bibfield  {journal} {\bibinfo  {journal} {Phys. Rev. E}\ }\textbf {\bibinfo
  {volume} {98}},\ \bibinfo {pages} {022409} (\bibinfo {year}
  {2018})}\BibitemShut {NoStop}%
\bibitem [{\citenamefont {Popović}\ \emph {et~al.}(2021)\citenamefont
  {Popović}, \citenamefont {Druelle}, \citenamefont {Dye}, \citenamefont
  {Jülicher},\ and\ \citenamefont {Wyart}}]{popovic21}%
  \BibitemOpen
  \bibfield  {author} {\bibinfo {author} {\bibfnamefont {M.}~\bibnamefont
  {Popović}}, \bibinfo {author} {\bibfnamefont {V.}~\bibnamefont {Druelle}},
  \bibinfo {author} {\bibfnamefont {N.~A.}\ \bibnamefont {Dye}}, \bibinfo
  {author} {\bibfnamefont {F.}~\bibnamefont {Jülicher}},\ and\ \bibinfo
  {author} {\bibfnamefont {M.}~\bibnamefont {Wyart}},\ }\bibfield  {title}
  {\bibinfo {title} {Inferring the flow properties of epithelial tissues from
  their geometry},\ }\href {https://doi.org/10.1088/1367-2630/abcbc7}
  {\bibfield  {journal} {\bibinfo  {journal} {New J. Phys.}\ }\textbf {\bibinfo
  {volume} {23}},\ \bibinfo {pages} {033004} (\bibinfo {year}
  {2021})}\BibitemShut {NoStop}%
\bibitem [{\citenamefont {Bi}\ \emph {et~al.}(2015)\citenamefont {Bi},
  \citenamefont {Lopez}, \citenamefont {Schwarz},\ and\ \citenamefont
  {Manning}}]{bi15}%
  \BibitemOpen
  \bibfield  {author} {\bibinfo {author} {\bibfnamefont {D.}~\bibnamefont
  {Bi}}, \bibinfo {author} {\bibfnamefont {J.~H.}\ \bibnamefont {Lopez}},
  \bibinfo {author} {\bibfnamefont {J.~M.}\ \bibnamefont {Schwarz}},\ and\
  \bibinfo {author} {\bibfnamefont {M.~L.}\ \bibnamefont {Manning}},\
  }\bibfield  {title} {\bibinfo {title} {A density-independent rigidity
  transition in biological tissues},\ }\href
  {https://doi.org/10.1038/nphys3471} {\bibfield  {journal} {\bibinfo
  {journal} {Nat. Phys.}\ }\textbf {\bibinfo {volume} {11}},\ \bibinfo {pages}
  {1074} (\bibinfo {year} {2015})}\BibitemShut {NoStop}%
\bibitem [{\citenamefont {Hannezo}\ and\ \citenamefont
  {Heisenberg}(2022)}]{hannezo22}%
  \BibitemOpen
  \bibfield  {author} {\bibinfo {author} {\bibfnamefont {E.}~\bibnamefont
  {Hannezo}}\ and\ \bibinfo {author} {\bibfnamefont {C.-P.}\ \bibnamefont
  {Heisenberg}},\ }\bibfield  {title} {\bibinfo {title} {Rigidity transitions
  in development and disease},\ }\href
  {https://doi.org/10.1016/j.tcb.2021.12.006} {\bibfield  {journal} {\bibinfo
  {journal} {Trends Cell Biol.}\ }\textbf {\bibinfo {volume} {32}},\ \bibinfo
  {pages} {433} (\bibinfo {year} {2022})}\BibitemShut {NoStop}%
\bibitem [{\citenamefont {Lenne}\ and\ \citenamefont
  {Trivedi}(2022)}]{lenne22}%
  \BibitemOpen
  \bibfield  {author} {\bibinfo {author} {\bibfnamefont {P.-F.}\ \bibnamefont
  {Lenne}}\ and\ \bibinfo {author} {\bibfnamefont {V.}~\bibnamefont
  {Trivedi}},\ }\bibfield  {title} {\bibinfo {title} {Sculpting tissues by
  phase transitions},\ }\href {https://doi.org/10.1038/s41467-022-28151-9}
  {\bibfield  {journal} {\bibinfo  {journal} {Nat. Commun.}\ }\textbf {\bibinfo
  {volume} {13}},\ \bibinfo {pages} {664} (\bibinfo {year} {2022})}\BibitemShut
  {NoStop}%
\bibitem [{\citenamefont {Mongera}\ \emph {et~al.}(2018)\citenamefont
  {Mongera}, \citenamefont {Rowghanian}, \citenamefont {Gustafson},
  \citenamefont {Shelton}, \citenamefont {Kealhofer}, \citenamefont {Carn},
  \citenamefont {Serwane}, \citenamefont {Lucio}, \citenamefont {Giammona},\
  and\ \citenamefont {Camp{\`a}s}}]{mongera18}%
  \BibitemOpen
  \bibfield  {author} {\bibinfo {author} {\bibfnamefont {A.}~\bibnamefont
  {Mongera}}, \bibinfo {author} {\bibfnamefont {P.}~\bibnamefont {Rowghanian}},
  \bibinfo {author} {\bibfnamefont {H.~J.}\ \bibnamefont {Gustafson}}, \bibinfo
  {author} {\bibfnamefont {E.}~\bibnamefont {Shelton}}, \bibinfo {author}
  {\bibfnamefont {D.~A.}\ \bibnamefont {Kealhofer}}, \bibinfo {author}
  {\bibfnamefont {E.~K.}\ \bibnamefont {Carn}}, \bibinfo {author}
  {\bibfnamefont {F.}~\bibnamefont {Serwane}}, \bibinfo {author} {\bibfnamefont
  {A.~A.}\ \bibnamefont {Lucio}}, \bibinfo {author} {\bibfnamefont
  {J.}~\bibnamefont {Giammona}},\ and\ \bibinfo {author} {\bibfnamefont
  {O.}~\bibnamefont {Camp{\`a}s}},\ }\bibfield  {title} {\bibinfo {title} {A
  fluid-to-solid jamming transition underlies vertebrate body axis
  elongation},\ }\href {https://doi.org/10.1038/s41586-018-0479-2} {\bibfield
  {journal} {\bibinfo  {journal} {Nature}\ }\textbf {\bibinfo {volume} {561}},\
  \bibinfo {pages} {401} (\bibinfo {year} {2018})}\BibitemShut {NoStop}%
\bibitem [{\citenamefont {Petridou}\ \emph {et~al.}(2019)\citenamefont
  {Petridou}, \citenamefont {Grigolon}, \citenamefont {Salbreux}, \citenamefont
  {Hannezo},\ and\ \citenamefont {Heisenberg}}]{petridou19}%
  \BibitemOpen
  \bibfield  {author} {\bibinfo {author} {\bibfnamefont {N.~I.}\ \bibnamefont
  {Petridou}}, \bibinfo {author} {\bibfnamefont {S.}~\bibnamefont {Grigolon}},
  \bibinfo {author} {\bibfnamefont {G.}~\bibnamefont {Salbreux}}, \bibinfo
  {author} {\bibfnamefont {E.}~\bibnamefont {Hannezo}},\ and\ \bibinfo {author}
  {\bibfnamefont {C.-P.}\ \bibnamefont {Heisenberg}},\ }\bibfield  {title}
  {\bibinfo {title} {Fluidization-mediated tissue spreading by mitotic cell
  rounding and non-canonical {Wnt} signalling},\ }\href
  {https://doi.org/10.1038/s41556-018-0247-4} {\bibfield  {journal} {\bibinfo
  {journal} {Nat. Cell Biol.}\ }\textbf {\bibinfo {volume} {21}},\ \bibinfo
  {pages} {169} (\bibinfo {year} {2019})}\BibitemShut {NoStop}%
\bibitem [{\citenamefont {Petridou}\ \emph {et~al.}(2021)\citenamefont
  {Petridou}, \citenamefont {Corominas-Murtra}, \citenamefont {Heisenberg},\
  and\ \citenamefont {Hannezo}}]{petridou21}%
  \BibitemOpen
  \bibfield  {author} {\bibinfo {author} {\bibfnamefont {N.~I.}\ \bibnamefont
  {Petridou}}, \bibinfo {author} {\bibfnamefont {B.}~\bibnamefont
  {Corominas-Murtra}}, \bibinfo {author} {\bibfnamefont {C.-P.}\ \bibnamefont
  {Heisenberg}},\ and\ \bibinfo {author} {\bibfnamefont {E.}~\bibnamefont
  {Hannezo}},\ }\bibfield  {title} {\bibinfo {title} {Rigidity percolation
  uncovers a structural basis for embryonic tissue phase transitions},\ }\href
  {https://doi.org/10.1016/j.cell.2021.02.017} {\bibfield  {journal} {\bibinfo
  {journal} {Cell}\ }\textbf {\bibinfo {volume} {184}},\ \bibinfo {pages}
  {1914} (\bibinfo {year} {2021})}\BibitemShut {NoStop}%
\bibitem [{\citenamefont {G{\'o}mez-G{\'a}lvez}\ \emph
  {et~al.}(2018)\citenamefont {G{\'o}mez-G{\'a}lvez}, \citenamefont
  {Vicente-Munuera}, \citenamefont {Tagua}, \citenamefont {Forja},
  \citenamefont {Castro}, \citenamefont {Letr{\'a}n}, \citenamefont
  {Valencia-Exp{\'o}sito}, \citenamefont {Grima}, \citenamefont
  {Berm{\'u}dez-Gallardo}, \citenamefont {Serrano-P{\'e}rez-Higueras},
  \citenamefont {Cavodeassi}, \citenamefont {Sotillos}, \citenamefont
  {Mart{\'i}n-Bermudo}, \citenamefont {M{\'a}rquez}, \citenamefont {Buceta},\
  and\ \citenamefont {Escudero}}]{gomezgalvez18}%
  \BibitemOpen
  \bibfield  {author} {\bibinfo {author} {\bibfnamefont {P.}~\bibnamefont
  {G{\'o}mez-G{\'a}lvez}}, \bibinfo {author} {\bibfnamefont {P.}~\bibnamefont
  {Vicente-Munuera}}, \bibinfo {author} {\bibfnamefont {A.}~\bibnamefont
  {Tagua}}, \bibinfo {author} {\bibfnamefont {C.}~\bibnamefont {Forja}},
  \bibinfo {author} {\bibfnamefont {A.~M.}\ \bibnamefont {Castro}}, \bibinfo
  {author} {\bibfnamefont {M.}~\bibnamefont {Letr{\'a}n}}, \bibinfo {author}
  {\bibfnamefont {A.}~\bibnamefont {Valencia-Exp{\'o}sito}}, \bibinfo {author}
  {\bibfnamefont {C.}~\bibnamefont {Grima}}, \bibinfo {author} {\bibfnamefont
  {M.}~\bibnamefont {Berm{\'u}dez-Gallardo}}, \bibinfo {author} {\bibfnamefont
  {{\'O}.}~\bibnamefont {Serrano-P{\'e}rez-Higueras}}, \bibinfo {author}
  {\bibfnamefont {F.}~\bibnamefont {Cavodeassi}}, \bibinfo {author}
  {\bibfnamefont {S.}~\bibnamefont {Sotillos}}, \bibinfo {author}
  {\bibfnamefont {M.~D.}\ \bibnamefont {Mart{\'i}n-Bermudo}}, \bibinfo {author}
  {\bibfnamefont {A.}~\bibnamefont {M{\'a}rquez}}, \bibinfo {author}
  {\bibfnamefont {J.}~\bibnamefont {Buceta}},\ and\ \bibinfo {author}
  {\bibfnamefont {L.~M.}\ \bibnamefont {Escudero}},\ }\bibfield  {title}
  {\bibinfo {title} {Scutoids are a geometrical solution to three-dimensional
  packing of epithelia},\ }\href {https://doi.org/10.1038/s41467-018-05376-1}
  {\bibfield  {journal} {\bibinfo  {journal} {Nat. Commun.}\ }\textbf {\bibinfo
  {volume} {9}},\ \bibinfo {pages} {2960} (\bibinfo {year} {2018})}\BibitemShut
  {NoStop}%
\bibitem [{\citenamefont {Gómez}\ \emph {et~al.}(2021)\citenamefont {Gómez},
  \citenamefont {Dumond}, \citenamefont {Hodel}, \citenamefont {Vetter},\ and\
  \citenamefont {Iber}}]{gomez21}%
  \BibitemOpen
  \bibfield  {author} {\bibinfo {author} {\bibfnamefont {H.~F.}\ \bibnamefont
  {Gómez}}, \bibinfo {author} {\bibfnamefont {M.~S.}\ \bibnamefont {Dumond}},
  \bibinfo {author} {\bibfnamefont {L.}~\bibnamefont {Hodel}}, \bibinfo
  {author} {\bibfnamefont {R.}~\bibnamefont {Vetter}},\ and\ \bibinfo {author}
  {\bibfnamefont {D.}~\bibnamefont {Iber}},\ }\bibfield  {title} {\bibinfo
  {title} {{3D} cell neighbour dynamics in growing pseudostratified
  epithelia},\ }\href {https://doi.org/10.7554/eLife.68135} {\bibfield
  {journal} {\bibinfo  {journal} {eLife}\ }\textbf {\bibinfo {volume} {10}},\
  \bibinfo {pages} {e68135} (\bibinfo {year} {2021})}\BibitemShut {NoStop}%
\bibitem [{\citenamefont {Lou}\ \emph {et~al.}(2023)\citenamefont {Lou},
  \citenamefont {Rupprecht}, \citenamefont {Theis}, \citenamefont {Hiraiwa},\
  and\ \citenamefont {Saunders}}]{lou23}%
  \BibitemOpen
  \bibfield  {author} {\bibinfo {author} {\bibfnamefont {Y.}~\bibnamefont
  {Lou}}, \bibinfo {author} {\bibfnamefont {J.-F.}\ \bibnamefont {Rupprecht}},
  \bibinfo {author} {\bibfnamefont {S.}~\bibnamefont {Theis}}, \bibinfo
  {author} {\bibfnamefont {T.}~\bibnamefont {Hiraiwa}},\ and\ \bibinfo {author}
  {\bibfnamefont {T.~E.}\ \bibnamefont {Saunders}},\ }\bibfield  {title}
  {\bibinfo {title} {Curvature-induced cell rearrangements in biological
  tissues},\ }\href {https://doi.org/10.1103/PhysRevLett.130.108401} {\bibfield
   {journal} {\bibinfo  {journal} {Phys. Rev. Lett.}\ }\textbf {\bibinfo
  {volume} {130}},\ \bibinfo {pages} {108401} (\bibinfo {year}
  {2023})}\BibitemShut {NoStop}%
\bibitem [{\citenamefont {Sarkar}\ and\ \citenamefont
  {Krajnc}(2024)}]{sarkar24}%
  \BibitemOpen
  \bibfield  {author} {\bibinfo {author} {\bibfnamefont {T.}~\bibnamefont
  {Sarkar}}\ and\ \bibinfo {author} {\bibfnamefont {M.}~\bibnamefont
  {Krajnc}},\ }\bibfield  {title} {\bibinfo {title} {Graph topological
  transformations in space-filling cell aggregates},\ }\href
  {https://doi.org/10.1371/journal.pcbi.1012089} {\bibfield  {journal}
  {\bibinfo  {journal} {PLoS Comp. Biol.}\ }\textbf {\bibinfo {volume} {20}},\
  \bibinfo {pages} {e1012089} (\bibinfo {year} {2024})}\BibitemShut {NoStop}%
\bibitem [{\citenamefont {Staddon}\ \emph {et~al.}(2019)\citenamefont
  {Staddon}, \citenamefont {Cavanaugh}, \citenamefont {Munro}, \citenamefont
  {Gardel},\ and\ \citenamefont {Banerjee}}]{staddon19}%
  \BibitemOpen
  \bibfield  {author} {\bibinfo {author} {\bibfnamefont {M.~F.}\ \bibnamefont
  {Staddon}}, \bibinfo {author} {\bibfnamefont {K.~E.}\ \bibnamefont
  {Cavanaugh}}, \bibinfo {author} {\bibfnamefont {E.~M.}\ \bibnamefont
  {Munro}}, \bibinfo {author} {\bibfnamefont {M.~L.}\ \bibnamefont {Gardel}},\
  and\ \bibinfo {author} {\bibfnamefont {S.}~\bibnamefont {Banerjee}},\
  }\bibfield  {title} {\bibinfo {title} {Mechanosensitive junction remodeling
  promotes robust epithelial morphogenesis},\ }\href
  {https://doi.org/10.1016/j.bpj.2019.09.027} {\bibfield  {journal} {\bibinfo
  {journal} {Biophys. J.}\ }\textbf {\bibinfo {volume} {117}},\ \bibinfo
  {pages} {1739} (\bibinfo {year} {2019})}\BibitemShut {NoStop}%
\bibitem [{\citenamefont {Yan}\ and\ \citenamefont {Bi}(2019)}]{yan19}%
  \BibitemOpen
  \bibfield  {author} {\bibinfo {author} {\bibfnamefont {L.}~\bibnamefont
  {Yan}}\ and\ \bibinfo {author} {\bibfnamefont {D.}~\bibnamefont {Bi}},\
  }\bibfield  {title} {\bibinfo {title} {Multicellular rosettes drive
  fluid-solid transition in epithelial tissues},\ }\href
  {https://doi.org/10.1103/PhysRevX.9.011029} {\bibfield  {journal} {\bibinfo
  {journal} {Phys. Rev. X}\ }\textbf {\bibinfo {volume} {9}},\ \bibinfo {pages}
  {011029} (\bibinfo {year} {2019})}\BibitemShut {NoStop}%
\bibitem [{\citenamefont {Krajnc}(2020)}]{krajnc20}%
  \BibitemOpen
  \bibfield  {author} {\bibinfo {author} {\bibfnamefont {M.}~\bibnamefont
  {Krajnc}},\ }\bibfield  {title} {\bibinfo {title} {Solid–fluid transition
  and cell sorting in epithelia with junctional tension fluctuations},\ }\href
  {https://doi.org/10.1039/C9SM02310K} {\bibfield  {journal} {\bibinfo
  {journal} {Soft Matter}\ }\textbf {\bibinfo {volume} {16}},\ \bibinfo {pages}
  {3209} (\bibinfo {year} {2020})}\BibitemShut {NoStop}%
\bibitem [{\citenamefont {Krajnc}\ \emph {et~al.}(2021)\citenamefont {Krajnc},
  \citenamefont {Stern},\ and\ \citenamefont {Zankoc}}]{krajnc21}%
  \BibitemOpen
  \bibfield  {author} {\bibinfo {author} {\bibfnamefont {M.}~\bibnamefont
  {Krajnc}}, \bibinfo {author} {\bibfnamefont {T.}~\bibnamefont {Stern}},\ and\
  \bibinfo {author} {\bibfnamefont {C.}~\bibnamefont {Zankoc}},\ }\bibfield
  {title} {\bibinfo {title} {Active instability and nonlinear dynamics of
  cell-cell junctions},\ }\href
  {https://doi.org/10.1103/PhysRevLett.127.198103} {\bibfield  {journal}
  {\bibinfo  {journal} {Phys. Rev. Lett.}\ }\textbf {\bibinfo {volume} {127}},\
  \bibinfo {pages} {198103} (\bibinfo {year} {2021})}\BibitemShut {NoStop}%
\bibitem [{\citenamefont {Duclut}\ \emph {et~al.}(2021)\citenamefont {Duclut},
  \citenamefont {Paijmans}, \citenamefont {Inamdar}, \citenamefont {Modes},\
  and\ \citenamefont {Jülicher}}]{duclut21}%
  \BibitemOpen
  \bibfield  {author} {\bibinfo {author} {\bibfnamefont {C.}~\bibnamefont
  {Duclut}}, \bibinfo {author} {\bibfnamefont {J.}~\bibnamefont {Paijmans}},
  \bibinfo {author} {\bibfnamefont {M.~M.}\ \bibnamefont {Inamdar}}, \bibinfo
  {author} {\bibfnamefont {C.~D.}\ \bibnamefont {Modes}},\ and\ \bibinfo
  {author} {\bibfnamefont {F.}~\bibnamefont {Jülicher}},\ }\bibfield  {title}
  {\bibinfo {title} {Nonlinear rheology of cellular networks},\ }\href
  {https://doi.org/10.1016/j.cdev.2021.203746} {\bibfield  {journal} {\bibinfo
  {journal} {Cells Dev.}\ }\textbf {\bibinfo {volume} {168}},\ \bibinfo {pages}
  {203746} (\bibinfo {year} {2021})}\BibitemShut {NoStop}%
\bibitem [{\citenamefont {Duclut}\ \emph {et~al.}(2022)\citenamefont {Duclut},
  \citenamefont {Paijmans}, \citenamefont {Inamdar}, \citenamefont {Modes},\
  and\ \citenamefont {J{\"u}licher}}]{duclut22}%
  \BibitemOpen
  \bibfield  {author} {\bibinfo {author} {\bibfnamefont {C.}~\bibnamefont
  {Duclut}}, \bibinfo {author} {\bibfnamefont {J.}~\bibnamefont {Paijmans}},
  \bibinfo {author} {\bibfnamefont {M.~M.}\ \bibnamefont {Inamdar}}, \bibinfo
  {author} {\bibfnamefont {C.~D.}\ \bibnamefont {Modes}},\ and\ \bibinfo
  {author} {\bibfnamefont {F.}~\bibnamefont {J{\"u}licher}},\ }\bibfield
  {title} {\bibinfo {title} {Active {T1} transitions in cellular networks},\
  }\href {https://doi.org/10.1140/epje/s10189-022-00175-5} {\bibfield
  {journal} {\bibinfo  {journal} {Eur. Phys. J. E}\ }\textbf {\bibinfo {volume}
  {45}},\ \bibinfo {pages} {29} (\bibinfo {year} {2022})}\BibitemShut {NoStop}%
\bibitem [{\citenamefont {Jain}\ \emph {et~al.}(2023)\citenamefont {Jain},
  \citenamefont {Voigt},\ and\ \citenamefont {Angheluta}}]{jain23}%
  \BibitemOpen
  \bibfield  {author} {\bibinfo {author} {\bibfnamefont {H.~P.}\ \bibnamefont
  {Jain}}, \bibinfo {author} {\bibfnamefont {A.}~\bibnamefont {Voigt}},\ and\
  \bibinfo {author} {\bibfnamefont {L.}~\bibnamefont {Angheluta}},\ }\bibfield
  {title} {\bibinfo {title} {Robust statistical properties of {T1} transitions
  in a multi-phase field model of cell monolayers},\ }\href
  {https://doi.org/10.1038/s41598-023-37064-6} {\bibfield  {journal} {\bibinfo
  {journal} {Sci. Rep.}\ }\textbf {\bibinfo {volume} {13}},\ \bibinfo {pages}
  {10096} (\bibinfo {year} {2023})}\BibitemShut {NoStop}%
\bibitem [{\citenamefont {Jain}\ \emph {et~al.}(2024)\citenamefont {Jain},
  \citenamefont {Voigt},\ and\ \citenamefont {Angheluta}}]{jain24}%
  \BibitemOpen
  \bibfield  {author} {\bibinfo {author} {\bibfnamefont {H.~P.}\ \bibnamefont
  {Jain}}, \bibinfo {author} {\bibfnamefont {A.}~\bibnamefont {Voigt}},\ and\
  \bibinfo {author} {\bibfnamefont {L.}~\bibnamefont {Angheluta}},\ }\bibfield
  {title} {\bibinfo {title} {From cell intercalation to flow, the importance of
  {T1} transitions},\ }\href {https://doi.org/10.1103/PhysRevResearch.6.033176}
  {\bibfield  {journal} {\bibinfo  {journal} {Phys. Rev. Res.}\ }\textbf
  {\bibinfo {volume} {6}},\ \bibinfo {pages} {033176} (\bibinfo {year}
  {2024})}\BibitemShut {NoStop}%
\bibitem [{\citenamefont {Staddon}\ and\ \citenamefont
  {Modes}(2025)}]{staddon25}%
  \BibitemOpen
  \bibfield  {author} {\bibinfo {author} {\bibfnamefont {M.~F.}\ \bibnamefont
  {Staddon}}\ and\ \bibinfo {author} {\bibfnamefont {C.~D.}\ \bibnamefont
  {Modes}},\ }\bibfield  {title} {\bibinfo {title} {Curved-edge vertex models
  and increased tissue fluidity},\ }\href
  {https://doi.org/10.1103/PhysRevResearch.7.013218} {\bibfield  {journal}
  {\bibinfo  {journal} {Phys. Rev. Res.}\ }\textbf {\bibinfo {volume} {7}},\
  \bibinfo {pages} {013218} (\bibinfo {year} {2025})}\BibitemShut {NoStop}%
\bibitem [{\citenamefont {De~Marzio}\ \emph {et~al.}(2025)\citenamefont
  {De~Marzio}, \citenamefont {Das}, \citenamefont {Fredberg},\ and\
  \citenamefont {Bi}}]{demarzio25}%
  \BibitemOpen
  \bibfield  {author} {\bibinfo {author} {\bibfnamefont {M.}~\bibnamefont
  {De~Marzio}}, \bibinfo {author} {\bibfnamefont {A.}~\bibnamefont {Das}},
  \bibinfo {author} {\bibfnamefont {J.~J.}\ \bibnamefont {Fredberg}},\ and\
  \bibinfo {author} {\bibfnamefont {D.}~\bibnamefont {Bi}},\ }\bibfield
  {title} {\bibinfo {title} {Epithelial layer fluidization by curvature-induced
  unjamming},\ }\href {https://doi.org/10.1103/PhysRevLett.134.138402}
  {\bibfield  {journal} {\bibinfo  {journal} {Phys. Rev. Lett.}\ }\textbf
  {\bibinfo {volume} {134}},\ \bibinfo {pages} {138402} (\bibinfo {year}
  {2025})}\BibitemShut {NoStop}%
\bibitem [{\citenamefont {Sknepnek}\ \emph {et~al.}(2023)\citenamefont
  {Sknepnek}, \citenamefont {Djafer-Cherif}, \citenamefont {Chuai},
  \citenamefont {Weijer},\ and\ \citenamefont {Henkes}}]{sknepnek23}%
  \BibitemOpen
  \bibfield  {author} {\bibinfo {author} {\bibfnamefont {R.}~\bibnamefont
  {Sknepnek}}, \bibinfo {author} {\bibfnamefont {I.}~\bibnamefont
  {Djafer-Cherif}}, \bibinfo {author} {\bibfnamefont {M.}~\bibnamefont
  {Chuai}}, \bibinfo {author} {\bibfnamefont {C.}~\bibnamefont {Weijer}},\ and\
  \bibinfo {author} {\bibfnamefont {S.}~\bibnamefont {Henkes}},\ }\bibfield
  {title} {\bibinfo {title} {Generating active {T1} transitions through
  mechanochemical feedback},\ }\href {https://doi.org/10.7554/eLife.79862}
  {\bibfield  {journal} {\bibinfo  {journal} {eLife}\ }\textbf {\bibinfo
  {volume} {12}},\ \bibinfo {pages} {e79862} (\bibinfo {year}
  {2023})}\BibitemShut {NoStop}%
\bibitem [{\citenamefont {Dye}\ \emph {et~al.}(2021)\citenamefont {Dye},
  \citenamefont {Popović}, \citenamefont {Iyer}, \citenamefont {Fuhrmann},
  \citenamefont {Piscitello-Gómez}, \citenamefont {Eaton},\ and\ \citenamefont
  {Jülicher}}]{dye21}%
  \BibitemOpen
  \bibfield  {author} {\bibinfo {author} {\bibfnamefont {N.~A.}\ \bibnamefont
  {Dye}}, \bibinfo {author} {\bibfnamefont {M.}~\bibnamefont {Popović}},
  \bibinfo {author} {\bibfnamefont {K.~V.}\ \bibnamefont {Iyer}}, \bibinfo
  {author} {\bibfnamefont {J.~F.}\ \bibnamefont {Fuhrmann}}, \bibinfo {author}
  {\bibfnamefont {R.}~\bibnamefont {Piscitello-Gómez}}, \bibinfo {author}
  {\bibfnamefont {S.}~\bibnamefont {Eaton}},\ and\ \bibinfo {author}
  {\bibfnamefont {F.}~\bibnamefont {Jülicher}},\ }\bibfield  {title} {\bibinfo
  {title} {Self-organized patterning of cell morphology via mechanosensitive
  feedback},\ }\href {https://doi.org/10.7554/eLife.57964} {\bibfield
  {journal} {\bibinfo  {journal} {eLife}\ }\textbf {\bibinfo {volume} {10}},\
  \bibinfo {pages} {e57964} (\bibinfo {year} {2021})}\BibitemShut {NoStop}%
\bibitem [{\citenamefont {Andreescu}\ and\ \citenamefont
  {Gelca}(2009)}]{andreescu}%
  \BibitemOpen
  \bibfield  {author} {\bibinfo {author} {\bibfnamefont {T.}~\bibnamefont
  {Andreescu}}\ and\ \bibinfo {author} {\bibfnamefont {R.}~\bibnamefont
  {Gelca}},\ }\href@noop {} {\emph {\bibinfo {title} {Mathematical Olympiad
  Challenges}}},\ \bibinfo {edition} {2nd}\ ed.\ (\bibinfo  {publisher}
  {Birkhäuser},\ \bibinfo {address} {Boston, MA},\ \bibinfo {year} {2009})\
  p.~\bibinfo {pages} {4}\BibitemShut {NoStop}%
\bibitem [{\citenamefont {Du}\ \emph {et~al.}(1987)\citenamefont {Du},
  \citenamefont {Hwang}, \citenamefont {Song},\ and\ \citenamefont
  {Ting}}]{du87}%
  \BibitemOpen
  \bibfield  {author} {\bibinfo {author} {\bibfnamefont {D.~Z.}\ \bibnamefont
  {Du}}, \bibinfo {author} {\bibfnamefont {F.~K.}\ \bibnamefont {Hwang}},
  \bibinfo {author} {\bibfnamefont {G.~D.}\ \bibnamefont {Song}},\ and\
  \bibinfo {author} {\bibfnamefont {G.~Y.}\ \bibnamefont {Ting}},\ }\bibfield
  {title} {\bibinfo {title} {Steiner minimal trees on sets of four points},\
  }\href {https://doi.org/10.1007/BF02187892} {\bibfield  {journal} {\bibinfo
  {journal} {Discrete Comput. Geom.}\ }\textbf {\bibinfo {volume} {2}},\
  \bibinfo {pages} {401} (\bibinfo {year} {1987})}\BibitemShut {NoStop}%
\bibitem [{\citenamefont {Gilbert}\ and\ \citenamefont
  {Pollak}(1968)}]{gilbert68}%
  \BibitemOpen
  \bibfield  {author} {\bibinfo {author} {\bibfnamefont {E.~N.}\ \bibnamefont
  {Gilbert}}\ and\ \bibinfo {author} {\bibfnamefont {H.~O.}\ \bibnamefont
  {Pollak}},\ }\bibfield  {title} {\bibinfo {title} {Steiner minimal trees},\
  }\href {https://doi.org/https://doi.org/10.1137/0116001} {\bibfield
  {journal} {\bibinfo  {journal} {SIAM J. Appl. Math.}\ }\textbf {\bibinfo
  {volume} {16}},\ \bibinfo {pages} {1} (\bibinfo {year} {1968})}\BibitemShut
  {NoStop}%
\bibitem [{\citenamefont {Weng}(1994)}]{weng94}%
  \BibitemOpen
  \bibfield  {author} {\bibinfo {author} {\bibfnamefont {J.~F.}\ \bibnamefont
  {Weng}},\ }\bibfield  {title} {\bibinfo {title} {Variational approach and
  {Steiner} minimal trees on four points},\ }\href
  {https://doi.org/10.1016/0012-365X(94)90244-5} {\bibfield  {journal}
  {\bibinfo  {journal} {Discrete Math.}\ }\textbf {\bibinfo {volume} {132}},\
  \bibinfo {pages} {349} (\bibinfo {year} {1994})}\BibitemShut {NoStop}%
\bibitem [{\citenamefont {Coxeter}(1969)}]{coxeter}%
  \BibitemOpen
  \bibfield  {author} {\bibinfo {author} {\bibfnamefont {H.~S.~M.}\
  \bibnamefont {Coxeter}},\ }\href@noop {} {\emph {\bibinfo {title}
  {Introduction to Geometry}}},\ \bibinfo {edition} {2nd}\ ed.\ (\bibinfo
  {publisher} {John Wiley \& Sons},\ \bibinfo {address} {New York, NY},\
  \bibinfo {year} {1969})\ \bibinfo {note} {{Chap.} 1.4, p. 11, and Chap. 1.8,
  pp. 20--23.}\BibitemShut {Stop}%
\bibitem [{\citenamefont {Johnson}(1960)}]{johnson}%
  \BibitemOpen
  \bibfield  {author} {\bibinfo {author} {\bibfnamefont {R.~A.}\ \bibnamefont
  {Johnson}},\ }\href@noop {} {\emph {\bibinfo {title} {Advanced {Euclidean}
  Geometry}}},\ \bibinfo {edition} {2nd}\ ed.\ (\bibinfo  {publisher} {Dover},\
  \bibinfo {address} {New York, NY},\ \bibinfo {year} {1960})\ \bibinfo {note}
  {{Par.} 92, pp. 62--64, Par.~97, pp. 68--69, Par.~275, p. 174, Par. 352, pp.
  218--220, Par.~355, 356, pp. 221--223.}\BibitemShut {Stop}%
\bibitem [{\citenamefont {Gueron}\ and\ \citenamefont
  {Tessler}(2002)}]{gueron02}%
  \BibitemOpen
  \bibfield  {author} {\bibinfo {author} {\bibfnamefont {S.}~\bibnamefont
  {Gueron}}\ and\ \bibinfo {author} {\bibfnamefont {R.}~\bibnamefont
  {Tessler}},\ }\bibfield  {title} {\bibinfo {title} {The {Fermat--Steiner}
  problem},\ }\href {http://www.jstor.org/stable/2695644?origin=JSTOR-pdf}
  {\bibfield  {journal} {\bibinfo  {journal} {Amer. Math. Monthly}\ }\textbf
  {\bibinfo {volume} {109}},\ \bibinfo {pages} {443} (\bibinfo {year}
  {2002})}\BibitemShut {NoStop}%
\bibitem [{\citenamefont {Saul}(2019)}]{saul}%
  \BibitemOpen
  \bibfield  {author} {\bibinfo {author} {\bibfnamefont {M.}~\bibnamefont
  {Saul}},\ }\href {https://doi.org/10.1090/mbk/070} {\emph {\bibinfo {title}
  {Hadamard's {Plane Geometry}: A Reader's Companion}}}\ (\bibinfo  {publisher}
  {American Mathematical Society},\ \bibinfo {address} {Providence, RI},\
  \bibinfo {year} {2019})\ p.~\bibinfo {pages} {7}\BibitemShut {NoStop}%
\bibitem [{Note1()}]{Note1}%
  \BibitemOpen
  \bibinfo {note} {See Supplemental Material at [url to be inserted], which
  includes Refs.~\cite
  {farhadifar07,staple10,fletcher14,fletcher16,alt17,grossman22,bi15,popovic21},
  for details of the calculations and numerical results for a mean-field vertex
  model of an isolated T1 transition.}\BibitemShut {Stop}%
\bibitem [{\citenamefont {Farhadifar}\ \emph {et~al.}(2007)\citenamefont
  {Farhadifar}, \citenamefont {Röper}, \citenamefont {Aigouy}, \citenamefont
  {Eaton},\ and\ \citenamefont {Jülicher}}]{farhadifar07}%
  \BibitemOpen
  \bibfield  {author} {\bibinfo {author} {\bibfnamefont {R.}~\bibnamefont
  {Farhadifar}}, \bibinfo {author} {\bibfnamefont {J.-C.}\ \bibnamefont
  {Röper}}, \bibinfo {author} {\bibfnamefont {B.}~\bibnamefont {Aigouy}},
  \bibinfo {author} {\bibfnamefont {S.}~\bibnamefont {Eaton}},\ and\ \bibinfo
  {author} {\bibfnamefont {F.}~\bibnamefont {Jülicher}},\ }\bibfield  {title}
  {\bibinfo {title} {The influence of cell mechanics, cell-cell interactions,
  and proliferation on epithelial packing},\ }\href
  {https://doi.org/10.1016/j.cub.2007.11.049} {\bibfield  {journal} {\bibinfo
  {journal} {Curr. Biol.}\ }\textbf {\bibinfo {volume} {17}},\ \bibinfo {pages}
  {2095} (\bibinfo {year} {2007})}\BibitemShut {NoStop}%
\bibitem [{\citenamefont {Staple}\ \emph {et~al.}(2010)\citenamefont {Staple},
  \citenamefont {Farhadifar}, \citenamefont {R{\"o}per}, \citenamefont
  {Aigouy}, \citenamefont {Eaton},\ and\ \citenamefont
  {J{\"u}licher}}]{staple10}%
  \BibitemOpen
  \bibfield  {author} {\bibinfo {author} {\bibfnamefont {D.~B.}\ \bibnamefont
  {Staple}}, \bibinfo {author} {\bibfnamefont {R.}~\bibnamefont {Farhadifar}},
  \bibinfo {author} {\bibfnamefont {J.-C.}\ \bibnamefont {R{\"o}per}}, \bibinfo
  {author} {\bibfnamefont {B.}~\bibnamefont {Aigouy}}, \bibinfo {author}
  {\bibfnamefont {S.}~\bibnamefont {Eaton}},\ and\ \bibinfo {author}
  {\bibfnamefont {F.}~\bibnamefont {J{\"u}licher}},\ }\bibfield  {title}
  {\bibinfo {title} {Mechanics and remodelling of cell packings in epithelia},\
  }\href {https://doi.org/10.1140/epje/i2010-10677-0} {\bibfield  {journal}
  {\bibinfo  {journal} {Eur. Phys. J. E}\ }\textbf {\bibinfo {volume} {33}},\
  \bibinfo {pages} {117} (\bibinfo {year} {2010})}\BibitemShut {NoStop}%
\bibitem [{\citenamefont {Fletcher}\ \emph {et~al.}(2014)\citenamefont
  {Fletcher}, \citenamefont {Osterfield}, \citenamefont {Baker},\ and\
  \citenamefont {Shvartsman}}]{fletcher14}%
  \BibitemOpen
  \bibfield  {author} {\bibinfo {author} {\bibfnamefont {A.~G.}\ \bibnamefont
  {Fletcher}}, \bibinfo {author} {\bibfnamefont {M.}~\bibnamefont
  {Osterfield}}, \bibinfo {author} {\bibfnamefont {R.~E.}\ \bibnamefont
  {Baker}},\ and\ \bibinfo {author} {\bibfnamefont {S.~Y.}\ \bibnamefont
  {Shvartsman}},\ }\bibfield  {title} {\bibinfo {title} {Vertex models of
  epithelial morphogenesis},\ }\href
  {https://doi.org/10.1016/j.bpj.2013.11.4498} {\bibfield  {journal} {\bibinfo
  {journal} {Biophys. J.}\ }\textbf {\bibinfo {volume} {106}},\ \bibinfo
  {pages} {2291} (\bibinfo {year} {2014})}\BibitemShut {NoStop}%
\bibitem [{\citenamefont {Fletcher}\ \emph {et~al.}(2016)\citenamefont
  {Fletcher}, \citenamefont {Cooper},\ and\ \citenamefont
  {Baker}}]{fletcher16}%
  \BibitemOpen
  \bibfield  {author} {\bibinfo {author} {\bibfnamefont {A.~G.}\ \bibnamefont
  {Fletcher}}, \bibinfo {author} {\bibfnamefont {F.}~\bibnamefont {Cooper}},\
  and\ \bibinfo {author} {\bibfnamefont {R.~E.}\ \bibnamefont {Baker}},\
  }\bibfield  {title} {\bibinfo {title} {Mechanocellular models of epithelial
  morphogenesis},\ }\href {https://doi.org/10.1098/rstb.2015.0519} {\bibfield
  {journal} {\bibinfo  {journal} {Phil. Trans. Roy. Soc. B}\ }\textbf {\bibinfo
  {volume} {372}},\ \bibinfo {pages} {20150519} (\bibinfo {year}
  {2016})}\BibitemShut {NoStop}%
\bibitem [{\citenamefont {Alt}\ \emph {et~al.}(2017)\citenamefont {Alt},
  \citenamefont {Ganguly},\ and\ \citenamefont {Salbreux}}]{alt17}%
  \BibitemOpen
  \bibfield  {author} {\bibinfo {author} {\bibfnamefont {S.}~\bibnamefont
  {Alt}}, \bibinfo {author} {\bibfnamefont {P.}~\bibnamefont {Ganguly}},\ and\
  \bibinfo {author} {\bibfnamefont {G.}~\bibnamefont {Salbreux}},\ }\bibfield
  {title} {\bibinfo {title} {Vertex models: from cell mechanics to tissue
  morphogenesis},\ }\href {https://doi.org/10.1098/rstb.2015.0520} {\bibfield
  {journal} {\bibinfo  {journal} {Phil. Trans. Roy. Soc. B}\ }\textbf {\bibinfo
  {volume} {372}},\ \bibinfo {pages} {20150520} (\bibinfo {year}
  {2017})}\BibitemShut {NoStop}%
\bibitem [{\citenamefont {Grossman}\ and\ \citenamefont
  {Joanny}(2022)}]{grossman22}%
  \BibitemOpen
  \bibfield  {author} {\bibinfo {author} {\bibfnamefont {D.}~\bibnamefont
  {Grossman}}\ and\ \bibinfo {author} {\bibfnamefont {J.-F.}\ \bibnamefont
  {Joanny}},\ }\bibfield  {title} {\bibinfo {title} {Instabilities and geometry
  of growing tissues},\ }\href {https://doi.org/10.1103/PhysRevLett.129.048102}
  {\bibfield  {journal} {\bibinfo  {journal} {Phys. Rev. Lett.}\ }\textbf
  {\bibinfo {volume} {129}},\ \bibinfo {pages} {048102} (\bibinfo {year}
  {2022})}\BibitemShut {NoStop}%
\bibitem [{\citenamefont {Church}(2019)}]{church19}%
  \BibitemOpen
  \bibfield  {author} {\bibinfo {author} {\bibfnamefont {R.~L.}\ \bibnamefont
  {Church}},\ }\bibinfo {title} {Understanding the {Weber} location paradigm},\
  in\ \href {https://doi.org/10.1007/978-3-030-19111-5_2} {\emph {\bibinfo
  {booktitle} {Contributions to Location Analysis}}},\ \bibinfo {editor}
  {edited by\ \bibinfo {editor} {\bibfnamefont {H.~A.}\ \bibnamefont {Eiselt}}\
  and\ \bibinfo {editor} {\bibfnamefont {V.}~\bibnamefont {Marianov}}}\
  (\bibinfo  {publisher} {Springer},\ \bibinfo {address} {Cham, Switzerland},\
  \bibinfo {year} {2019})\ pp.\ \bibinfo {pages} {69--88}\BibitemShut {NoStop}%
\bibitem [{\citenamefont {Church}\ \emph {et~al.}(2022)\citenamefont {Church},
  \citenamefont {Drezner},\ and\ \citenamefont {Tamir}}]{church22}%
  \BibitemOpen
  \bibfield  {author} {\bibinfo {author} {\bibfnamefont {R.~L.}\ \bibnamefont
  {Church}}, \bibinfo {author} {\bibfnamefont {Z.}~\bibnamefont {Drezner}},\
  and\ \bibinfo {author} {\bibfnamefont {A.}~\bibnamefont {Tamir}},\ }\bibfield
   {title} {\bibinfo {title} {Extensions to the {Weber} problem},\ }\href
  {https://doi.org/10.1016/j.cor.2022.105786} {\bibfield  {journal} {\bibinfo
  {journal} {Comput. Oper. Res.}\ }\textbf {\bibinfo {volume} {143}},\ \bibinfo
  {pages} {105786} (\bibinfo {year} {2022})}\BibitemShut {NoStop}%
\bibitem [{\citenamefont {Kirby}(1980)}]{kirby80}%
  \BibitemOpen
  \bibfield  {author} {\bibinfo {author} {\bibfnamefont {D.}~\bibnamefont
  {Kirby}},\ }\bibfield  {title} {\bibinfo {title} {Concurrences for a
  triangle},\ }\href {http://www.jstor.org/stable/2320382} {\bibfield
  {journal} {\bibinfo  {journal} {Amer. Math. Monthly}\ }\textbf {\bibinfo
  {volume} {87}},\ \bibinfo {pages} {45} (\bibinfo {year} {1980})}\BibitemShut
  {NoStop}%
\bibitem [{\citenamefont {Gale}(1996)}]{gale96}%
  \BibitemOpen
  \bibfield  {author} {\bibinfo {author} {\bibfnamefont {D.}~\bibnamefont
  {Gale}},\ }\bibfield  {title} {\bibinfo {title} {Mathematical
  entertainments},\ }\href {https://doi.org/10.1007/BF03024813} {\bibfield
  {journal} {\bibinfo  {journal} {Math. Intelligencer}\ }\textbf {\bibinfo
  {volume} {18}},\ \bibinfo {pages} {31} (\bibinfo {year} {1996})}\BibitemShut
  {NoStop}%
\bibitem [{\citenamefont {Vickers}(2015)}]{vickers15}%
  \BibitemOpen
  \bibfield  {author} {\bibinfo {author} {\bibfnamefont {G.~T.}\ \bibnamefont
  {Vickers}},\ }\bibfield  {title} {\bibinfo {title} {Reciprocal {Jacobi}
  triangles and the {McCay} cubic},\ }\href
  {https://forumgeom.fau.edu/FG2015volume15/FG201518.pdf} {\bibfield  {journal}
  {\bibinfo  {journal} {Forum Geom.}\ }\textbf {\bibinfo {volume} {15}},\
  \bibinfo {pages} {179} (\bibinfo {year} {2015})}\BibitemShut {NoStop}%
\bibitem [{\citenamefont {Curran}\ \emph {et~al.}(2017)\citenamefont {Curran},
  \citenamefont {Strandkvist}, \citenamefont {Bathmann}, \citenamefont {{de
  Gennes}}, \citenamefont {Kabla}, \citenamefont {Salbreux},\ and\
  \citenamefont {Baum}}]{curran17}%
  \BibitemOpen
  \bibfield  {author} {\bibinfo {author} {\bibfnamefont {S.}~\bibnamefont
  {Curran}}, \bibinfo {author} {\bibfnamefont {C.}~\bibnamefont {Strandkvist}},
  \bibinfo {author} {\bibfnamefont {J.}~\bibnamefont {Bathmann}}, \bibinfo
  {author} {\bibfnamefont {M.}~\bibnamefont {{de Gennes}}}, \bibinfo {author}
  {\bibfnamefont {A.}~\bibnamefont {Kabla}}, \bibinfo {author} {\bibfnamefont
  {G.}~\bibnamefont {Salbreux}},\ and\ \bibinfo {author} {\bibfnamefont
  {B.}~\bibnamefont {Baum}},\ }\bibfield  {title} {\bibinfo {title} {Myosin
  {II} controls junction fluctuations to guide epithelial tissue ordering},\
  }\href {https://doi.org/10.1016/j.devcel.2017.09.018} {\bibfield  {journal}
  {\bibinfo  {journal} {Dev. Cell}\ }\textbf {\bibinfo {volume} {43}},\
  \bibinfo {pages} {480} (\bibinfo {year} {2017})}\BibitemShut {NoStop}%
\bibitem [{\citenamefont {Tetley}\ \emph {et~al.}(2019)\citenamefont {Tetley},
  \citenamefont {Staddon}, \citenamefont {Heller}, \citenamefont {Hoppe},
  \citenamefont {Banerjee},\ and\ \citenamefont {Mao}}]{tetley19}%
  \BibitemOpen
  \bibfield  {author} {\bibinfo {author} {\bibfnamefont {R.~J.}\ \bibnamefont
  {Tetley}}, \bibinfo {author} {\bibfnamefont {M.~F.}\ \bibnamefont {Staddon}},
  \bibinfo {author} {\bibfnamefont {D.}~\bibnamefont {Heller}}, \bibinfo
  {author} {\bibfnamefont {A.}~\bibnamefont {Hoppe}}, \bibinfo {author}
  {\bibfnamefont {S.}~\bibnamefont {Banerjee}},\ and\ \bibinfo {author}
  {\bibfnamefont {Y.}~\bibnamefont {Mao}},\ }\bibfield  {title} {\bibinfo
  {title} {Tissue fluidity promotes epithelial wound healing},\ }\href
  {https://doi.org/10.1038/s41567-019-0618-1} {\bibfield  {journal} {\bibinfo
  {journal} {Nat. Phys.}\ }\textbf {\bibinfo {volume} {15}},\ \bibinfo {pages}
  {1195} (\bibinfo {year} {2019})}\BibitemShut {NoStop}%
\bibitem [{\citenamefont {Kim}\ \emph {et~al.}(2021)\citenamefont {Kim},
  \citenamefont {Pochitaloff}, \citenamefont {Stooke-Vaughan},\ and\
  \citenamefont {Camp{\`a}s}}]{kim21}%
  \BibitemOpen
  \bibfield  {author} {\bibinfo {author} {\bibfnamefont {S.}~\bibnamefont
  {Kim}}, \bibinfo {author} {\bibfnamefont {M.}~\bibnamefont {Pochitaloff}},
  \bibinfo {author} {\bibfnamefont {G.~A.}\ \bibnamefont {Stooke-Vaughan}},\
  and\ \bibinfo {author} {\bibfnamefont {O.}~\bibnamefont {Camp{\`a}s}},\
  }\bibfield  {title} {\bibinfo {title} {Embryonic tissues as active foams},\
  }\href {https://doi.org/10.1038/s41567-021-01215-1} {\bibfield  {journal}
  {\bibinfo  {journal} {Nat. Phys.}\ }\textbf {\bibinfo {volume} {17}},\
  \bibinfo {pages} {859} (\bibinfo {year} {2021})}\BibitemShut {NoStop}%
\bibitem [{\citenamefont {Yamamoto}\ \emph {et~al.}(2022)\citenamefont
  {Yamamoto}, \citenamefont {Sussman}, \citenamefont {Shibata},\ and\
  \citenamefont {Manning}}]{yamamoto22}%
  \BibitemOpen
  \bibfield  {author} {\bibinfo {author} {\bibfnamefont {T.}~\bibnamefont
  {Yamamoto}}, \bibinfo {author} {\bibfnamefont {D.~M.}\ \bibnamefont
  {Sussman}}, \bibinfo {author} {\bibfnamefont {T.}~\bibnamefont {Shibata}},\
  and\ \bibinfo {author} {\bibfnamefont {M.~L.}\ \bibnamefont {Manning}},\
  }\bibfield  {title} {\bibinfo {title} {Non-monotonic fluidization generated
  by fluctuating edge tensions in confluent tissues},\ }\href
  {https://doi.org/10.1039/D0SM01559H} {\bibfield  {journal} {\bibinfo
  {journal} {Soft Matter}\ }\textbf {\bibinfo {volume} {18}},\ \bibinfo {pages}
  {2168} (\bibinfo {year} {2022})}\BibitemShut {NoStop}%
\bibitem [{\citenamefont {Damavandi}\ \emph {et~al.}(2025)\citenamefont
  {Damavandi}, \citenamefont {Arzash}, \citenamefont {Lawson-Keister},\ and\
  \citenamefont {Manning}}]{damavandi25}%
  \BibitemOpen
  \bibfield  {author} {\bibinfo {author} {\bibfnamefont {O.~K.}\ \bibnamefont
  {Damavandi}}, \bibinfo {author} {\bibfnamefont {S.}~\bibnamefont {Arzash}},
  \bibinfo {author} {\bibfnamefont {E.}~\bibnamefont {Lawson-Keister}},\ and\
  \bibinfo {author} {\bibfnamefont {M.~L.}\ \bibnamefont {Manning}},\
  }\bibfield  {title} {\bibinfo {title} {Universality in the mechanical
  behavior of vertex models for biological tissues},\ }\href
  {https://doi.org/10.1103/9ktk-6rqc} {\bibfield  {journal} {\bibinfo
  {journal} {PRX Life}\ }\textbf {\bibinfo {volume} {3}},\ \bibinfo {pages}
  {033001} (\bibinfo {year} {2025})}\BibitemShut {NoStop}%
\bibitem [{\citenamefont {Dunham}(2000)}]{dunham}%
  \BibitemOpen
  \bibfield  {author} {\bibinfo {author} {\bibfnamefont {W.}~\bibnamefont
  {Dunham}},\ }\bibfield  {title} {\bibinfo {title} {Quadrilaterally
  speaking},\ }\href {http://www.jstor.org/stable/25678249} {\bibfield
  {journal} {\bibinfo  {journal} {Math. Horizons}\ }\textbf {\bibinfo {volume}
  {7}},\ \bibinfo {pages} {12} (\bibinfo {year} {2000})}\BibitemShut {NoStop}%
\bibitem [{Note2()}]{Note2}%
  \BibitemOpen
  \bibinfo {note} {This fact is surely folklore, but I could not find an exact
  reference. To prove it, I denote by $\protect \boldsymbol {r_A},\protect
  \boldsymbol {r_B},\protect \boldsymbol {r_C},\protect \boldsymbol {r_D}$ the
  position vectors of $A,B,C,D$ and compute the gradient of $E(\protect
  \boldsymbol {r})=\|\protect \boldsymbol {r}-\protect \boldsymbol
  {r_A}\|^2+\|\protect \boldsymbol {r}-\protect \boldsymbol
  {r_B}\|^2+\|\protect \boldsymbol {r}-\protect \boldsymbol
  {r_C}\|^2+\|\protect \boldsymbol {r}-\protect \boldsymbol {r_D}\|^2$,
  $\protect \boldsymbol {\nabla }E=2\protect \bigl (4\protect \boldsymbol
  {r}-\protect \boldsymbol {r_A}-\protect \boldsymbol {r_B}-\protect
  \boldsymbol {r_C}-\protect \boldsymbol {r_D}\protect \bigr )$, which has a
  single stationary point at $\protect \boldsymbol {r}=(\protect \boldsymbol
  {r_A}+\protect \boldsymbol {r_B}+\protect \boldsymbol {r_C}+\protect
  \boldsymbol {r_D})/4$, i.e., at the barycentre, and this is clearly a
  minimum.}\BibitemShut {Stop}%
\bibitem [{\citenamefont {Sahu}\ \emph {et~al.}(2020)\citenamefont {Sahu},
  \citenamefont {Kang}, \citenamefont {Erdemci-Tandogan},\ and\ \citenamefont
  {Manning}}]{sahu20}%
  \BibitemOpen
  \bibfield  {author} {\bibinfo {author} {\bibfnamefont {P.}~\bibnamefont
  {Sahu}}, \bibinfo {author} {\bibfnamefont {J.}~\bibnamefont {Kang}}, \bibinfo
  {author} {\bibfnamefont {G.}~\bibnamefont {Erdemci-Tandogan}},\ and\ \bibinfo
  {author} {\bibfnamefont {M.~L.}\ \bibnamefont {Manning}},\ }\bibfield
  {title} {\bibinfo {title} {Linear and nonlinear mechanical responses can be
  quite different in models for biological tissues},\ }\href
  {https://doi.org/10.1039/C9SM01068H} {\bibfield  {journal} {\bibinfo
  {journal} {Soft Matter}\ }\textbf {\bibinfo {volume} {16}},\ \bibinfo {pages}
  {1850} (\bibinfo {year} {2020})}\BibitemShut {NoStop}%
\bibitem [{\citenamefont {Murisic}\ \emph {et~al.}(2015)\citenamefont
  {Murisic}, \citenamefont {Hakim}, \citenamefont {Kevrekidis}, \citenamefont
  {Shvartsman},\ and\ \citenamefont {Audoly}}]{murisic15}%
  \BibitemOpen
  \bibfield  {author} {\bibinfo {author} {\bibfnamefont {N.}~\bibnamefont
  {Murisic}}, \bibinfo {author} {\bibfnamefont {V.}~\bibnamefont {Hakim}},
  \bibinfo {author} {\bibfnamefont {I.~G.}\ \bibnamefont {Kevrekidis}},
  \bibinfo {author} {\bibfnamefont {S.~Y.}\ \bibnamefont {Shvartsman}},\ and\
  \bibinfo {author} {\bibfnamefont {B.}~\bibnamefont {Audoly}},\ }\bibfield
  {title} {\bibinfo {title} {From discrete to continuum models of
  three-dimensional deformations in epithelial sheets},\ }\href
  {https://doi.org/10.1016/j.bpj.2015.05.019} {\bibfield  {journal} {\bibinfo
  {journal} {Biophys. J.}\ }\textbf {\bibinfo {volume} {109}},\ \bibinfo
  {pages} {154} (\bibinfo {year} {2015})}\BibitemShut {NoStop}%
\bibitem [{\citenamefont {Nestor-Bergmann}\ \emph {et~al.}(2017)\citenamefont
  {Nestor-Bergmann}, \citenamefont {Goddard}, \citenamefont {Woolner},\ and\
  \citenamefont {Jensen}}]{nestorbergmann17}%
  \BibitemOpen
  \bibfield  {author} {\bibinfo {author} {\bibfnamefont {A.}~\bibnamefont
  {Nestor-Bergmann}}, \bibinfo {author} {\bibfnamefont {G.}~\bibnamefont
  {Goddard}}, \bibinfo {author} {\bibfnamefont {S.}~\bibnamefont {Woolner}},\
  and\ \bibinfo {author} {\bibfnamefont {O.~E.}\ \bibnamefont {Jensen}},\
  }\bibfield  {title} {\bibinfo {title} {Relating cell shape and mechanical
  stress in a spatially disordered epithelium using a vertex-based model},\
  }\href {https://doi.org/10.1093/imammb/dqx008} {\bibfield  {journal}
  {\bibinfo  {journal} {Math. Med. Biol.}\ }\textbf {\bibinfo {volume} {35}},\
  \bibinfo {pages} {i1} (\bibinfo {year} {2017})}\BibitemShut {NoStop}%
\bibitem [{\citenamefont {Tong}\ \emph {et~al.}(2022)\citenamefont {Tong},
  \citenamefont {Singh}, \citenamefont {Sknepnek},\ and\ \citenamefont
  {Košmrlj}}]{tong22}%
  \BibitemOpen
  \bibfield  {author} {\bibinfo {author} {\bibfnamefont {S.}~\bibnamefont
  {Tong}}, \bibinfo {author} {\bibfnamefont {N.~K.}\ \bibnamefont {Singh}},
  \bibinfo {author} {\bibfnamefont {R.}~\bibnamefont {Sknepnek}},\ and\
  \bibinfo {author} {\bibfnamefont {A.}~\bibnamefont {Košmrlj}},\ }\bibfield
  {title} {\bibinfo {title} {Linear viscoelastic properties of the vertex model
  for epithelial tissues},\ }\href
  {https://doi.org/10.1371/journal.pcbi.1010135} {\bibfield  {journal}
  {\bibinfo  {journal} {PLoS Comput. Biol.}\ }\textbf {\bibinfo {volume}
  {18}},\ \bibinfo {pages} {e1010135} (\bibinfo {year} {2022})}\BibitemShut
  {NoStop}%
\bibitem [{\citenamefont {Hernandez}\ \emph {et~al.}(2022)\citenamefont
  {Hernandez}, \citenamefont {Staddon}, \citenamefont {Bowick}, \citenamefont
  {Marchetti},\ and\ \citenamefont {Moshe}}]{hernandez22}%
  \BibitemOpen
  \bibfield  {author} {\bibinfo {author} {\bibfnamefont {A.}~\bibnamefont
  {Hernandez}}, \bibinfo {author} {\bibfnamefont {M.~F.}\ \bibnamefont
  {Staddon}}, \bibinfo {author} {\bibfnamefont {M.~J.}\ \bibnamefont {Bowick}},
  \bibinfo {author} {\bibfnamefont {M.~C.}\ \bibnamefont {Marchetti}},\ and\
  \bibinfo {author} {\bibfnamefont {M.}~\bibnamefont {Moshe}},\ }\bibfield
  {title} {\bibinfo {title} {Anomalous elasticity of a cellular tissue vertex
  model},\ }\href {https://doi.org/10.1103/PhysRevE.105.064611} {\bibfield
  {journal} {\bibinfo  {journal} {Phys. Rev. E}\ }\textbf {\bibinfo {volume}
  {105}},\ \bibinfo {pages} {064611} (\bibinfo {year} {2022})}\BibitemShut
  {NoStop}%
\bibitem [{\citenamefont {Tong}\ \emph {et~al.}(2023)\citenamefont {Tong},
  \citenamefont {Sknepnek},\ and\ \citenamefont {Košmrlj}}]{tong23}%
  \BibitemOpen
  \bibfield  {author} {\bibinfo {author} {\bibfnamefont {S.}~\bibnamefont
  {Tong}}, \bibinfo {author} {\bibfnamefont {R.}~\bibnamefont {Sknepnek}},\
  and\ \bibinfo {author} {\bibfnamefont {A.}~\bibnamefont {Košmrlj}},\
  }\bibfield  {title} {\bibinfo {title} {Linear viscoelastic response of the
  vertex model with internal and external dissipation: Normal modes analysis},\
  }\href {https://doi.org/10.1103/PhysRevResearch.5.013143} {\bibfield
  {journal} {\bibinfo  {journal} {Phys. Rev. Res.}\ }\textbf {\bibinfo {volume}
  {5}},\ \bibinfo {pages} {013143} (\bibinfo {year} {2023})}\BibitemShut
  {NoStop}%
\bibitem [{\citenamefont {Staddon}\ \emph {et~al.}(2023)\citenamefont
  {Staddon}, \citenamefont {Hernandez}, \citenamefont {Bowick}, \citenamefont
  {Moshe},\ and\ \citenamefont {Marchetti}}]{staddon23}%
  \BibitemOpen
  \bibfield  {author} {\bibinfo {author} {\bibfnamefont {M.~F.}\ \bibnamefont
  {Staddon}}, \bibinfo {author} {\bibfnamefont {A.}~\bibnamefont {Hernandez}},
  \bibinfo {author} {\bibfnamefont {M.~J.}\ \bibnamefont {Bowick}}, \bibinfo
  {author} {\bibfnamefont {M.}~\bibnamefont {Moshe}},\ and\ \bibinfo {author}
  {\bibfnamefont {M.~C.}\ \bibnamefont {Marchetti}},\ }\bibfield  {title}
  {\bibinfo {title} {The role of non-affine deformations in the elastic
  behavior of the cellular vertex model},\ }\href
  {https://doi.org/10.1039/D2SM01580C} {\bibfield  {journal} {\bibinfo
  {journal} {Soft Matter}\ }\textbf {\bibinfo {volume} {19}},\ \bibinfo {pages}
  {3080} (\bibinfo {year} {2023})}\BibitemShut {NoStop}%
\bibitem [{\citenamefont {Hernandez}\ \emph {et~al.}(2023)\citenamefont
  {Hernandez}, \citenamefont {Staddon}, \citenamefont {Moshe},\ and\
  \citenamefont {Marchetti}}]{hernandez23}%
  \BibitemOpen
  \bibfield  {author} {\bibinfo {author} {\bibfnamefont {A.}~\bibnamefont
  {Hernandez}}, \bibinfo {author} {\bibfnamefont {M.~F.}\ \bibnamefont
  {Staddon}}, \bibinfo {author} {\bibfnamefont {M.}~\bibnamefont {Moshe}},\
  and\ \bibinfo {author} {\bibfnamefont {M.~C.}\ \bibnamefont {Marchetti}},\
  }\bibfield  {title} {\bibinfo {title} {Finite elasticity of the vertex model
  and its role in rigidity of curved cellular tissues},\ }\href
  {https://doi.org/10.1039/D3SM00874F} {\bibfield  {journal} {\bibinfo
  {journal} {Soft Matter}\ }\textbf {\bibinfo {volume} {19}},\ \bibinfo {pages}
  {7744} (\bibinfo {year} {2023})}\BibitemShut {NoStop}%
\bibitem [{\citenamefont {Kim}\ \emph {et~al.}(2024)\citenamefont {Kim},
  \citenamefont {Zhang},\ and\ \citenamefont {Schwarz}}]{kim24}%
  \BibitemOpen
  \bibfield  {author} {\bibinfo {author} {\bibfnamefont {K.}~\bibnamefont
  {Kim}}, \bibinfo {author} {\bibfnamefont {T.}~\bibnamefont {Zhang}},\ and\
  \bibinfo {author} {\bibfnamefont {J.~M.}\ \bibnamefont {Schwarz}},\
  }\bibfield  {title} {\bibinfo {title} {Mean-field elastic moduli of a
  three-dimensional, cell-based vertex model},\ }\href
  {https://doi.org/10.1088/1367-2630/ad3099} {\bibfield  {journal} {\bibinfo
  {journal} {New J. Phys.}\ }\textbf {\bibinfo {volume} {26}},\ \bibinfo
  {pages} {043009} (\bibinfo {year} {2024})}\BibitemShut {NoStop}%
\bibitem [{\citenamefont {Fielding}\ \emph {et~al.}(2023)\citenamefont
  {Fielding}, \citenamefont {Cochran}, \citenamefont {Huang}, \citenamefont
  {Bi},\ and\ \citenamefont {Marchetti}}]{fielding23}%
  \BibitemOpen
  \bibfield  {author} {\bibinfo {author} {\bibfnamefont {S.~M.}\ \bibnamefont
  {Fielding}}, \bibinfo {author} {\bibfnamefont {J.~O.}\ \bibnamefont
  {Cochran}}, \bibinfo {author} {\bibfnamefont {J.}~\bibnamefont {Huang}},
  \bibinfo {author} {\bibfnamefont {D.}~\bibnamefont {Bi}},\ and\ \bibinfo
  {author} {\bibfnamefont {M.~C.}\ \bibnamefont {Marchetti}},\ }\bibfield
  {title} {\bibinfo {title} {Constitutive model for the rheology of biological
  tissue},\ }\href {https://doi.org/10.1103/PhysRevE.108.L042602} {\bibfield
  {journal} {\bibinfo  {journal} {Phys. Rev. E}\ }\textbf {\bibinfo {volume}
  {108}},\ \bibinfo {pages} {L042602} (\bibinfo {year} {2023})}\BibitemShut
  {NoStop}%
\bibitem [{\citenamefont {Pérez-Verdugo}\ and\ \citenamefont
  {Soto}(2023)}]{perez23}%
  \BibitemOpen
  \bibfield  {author} {\bibinfo {author} {\bibfnamefont {F.}~\bibnamefont
  {Pérez-Verdugo}}\ and\ \bibinfo {author} {\bibfnamefont {R.}~\bibnamefont
  {Soto}},\ }\bibfield  {title} {\bibinfo {title} {Continuum description of
  confluent tissues with spatial heterogeneous activity},\ }\href
  {https://doi.org/10.1039/D3SM00254C} {\bibfield  {journal} {\bibinfo
  {journal} {Soft Matter}\ }\textbf {\bibinfo {volume} {19}},\ \bibinfo {pages}
  {6501} (\bibinfo {year} {2023})}\BibitemShut {NoStop}%
\bibitem [{\citenamefont {Triguero-Platero}\ \emph {et~al.}(2023)\citenamefont
  {Triguero-Platero}, \citenamefont {Ziebert},\ and\ \citenamefont
  {Bonilla}}]{triguero23}%
  \BibitemOpen
  \bibfield  {author} {\bibinfo {author} {\bibfnamefont {G.}~\bibnamefont
  {Triguero-Platero}}, \bibinfo {author} {\bibfnamefont {F.}~\bibnamefont
  {Ziebert}},\ and\ \bibinfo {author} {\bibfnamefont {L.~L.}\ \bibnamefont
  {Bonilla}},\ }\bibfield  {title} {\bibinfo {title} {Coarse-graining the
  vertex model and its response to shear},\ }\href
  {https://doi.org/10.1103/PhysRevE.108.044118} {\bibfield  {journal} {\bibinfo
   {journal} {Phys. Rev. E}\ }\textbf {\bibinfo {volume} {108}},\ \bibinfo
  {pages} {044118} (\bibinfo {year} {2023})}\BibitemShut {NoStop}%
\bibitem [{\citenamefont {Grossman}\ and\ \citenamefont
  {Joanny}(2025)}]{grossman25}%
  \BibitemOpen
  \bibfield  {author} {\bibinfo {author} {\bibfnamefont {D.}~\bibnamefont
  {Grossman}}\ and\ \bibinfo {author} {\bibfnamefont {J.-F.}\ \bibnamefont
  {Joanny}},\ }\bibfield  {title} {\bibinfo {title} {Rheology of the vertex
  model of tissues: Simple shear and oscillatory geometries},\ }\href
  {https://doi.org/10.1103/PhysRevResearch.7.013039} {\bibfield  {journal}
  {\bibinfo  {journal} {Phys. Rev. Res.}\ }\textbf {\bibinfo {volume} {7}},\
  \bibinfo {pages} {013039} (\bibinfo {year} {2025})}\BibitemShut {NoStop}%
\bibitem [{\citenamefont {Krajnc}\ and\ \citenamefont
  {Ziherl}(2015)}]{krajnc15}%
  \BibitemOpen
  \bibfield  {author} {\bibinfo {author} {\bibfnamefont {M.}~\bibnamefont
  {Krajnc}}\ and\ \bibinfo {author} {\bibfnamefont {P.}~\bibnamefont
  {Ziherl}},\ }\bibfield  {title} {\bibinfo {title} {Theory of epithelial
  elasticity},\ }\href {https://doi.org/10.1103/PhysRevE.92.052713} {\bibfield
  {journal} {\bibinfo  {journal} {Phys. Rev. E}\ }\textbf {\bibinfo {volume}
  {92}},\ \bibinfo {pages} {052713} (\bibinfo {year} {2015})}\BibitemShut
  {NoStop}%
\bibitem [{\citenamefont {Haas}\ and\ \citenamefont
  {Goldstein}(2019)}]{haas19}%
  \BibitemOpen
  \bibfield  {author} {\bibinfo {author} {\bibfnamefont {P.~A.}\ \bibnamefont
  {Haas}}\ and\ \bibinfo {author} {\bibfnamefont {R.~E.}\ \bibnamefont
  {Goldstein}},\ }\bibfield  {title} {\bibinfo {title} {Nonlinear and nonlocal
  elasticity in coarse-grained differential-tension models of epithelia},\
  }\href {https://doi.org/10.1103/PhysRevE.99.022411} {\bibfield  {journal}
  {\bibinfo  {journal} {Phys. Rev. E}\ }\textbf {\bibinfo {volume} {99}},\
  \bibinfo {pages} {022411} (\bibinfo {year} {2019})}\BibitemShut {NoStop}%
\bibitem [{\citenamefont {Andren\v{s}ek}\ \emph {et~al.}(2023)\citenamefont
  {Andren\v{s}ek}, \citenamefont {Ziherl},\ and\ \citenamefont
  {Krajnc}}]{andrensek23}%
  \BibitemOpen
  \bibfield  {author} {\bibinfo {author} {\bibfnamefont {U.}~\bibnamefont
  {Andren\v{s}ek}}, \bibinfo {author} {\bibfnamefont {P.}~\bibnamefont
  {Ziherl}},\ and\ \bibinfo {author} {\bibfnamefont {M.}~\bibnamefont
  {Krajnc}},\ }\bibfield  {title} {\bibinfo {title} {Wrinkling instability in
  unsupported epithelial sheets},\ }\href
  {https://doi.org/10.1103/PhysRevLett.130.198401} {\bibfield  {journal}
  {\bibinfo  {journal} {Phys. Rev. Lett.}\ }\textbf {\bibinfo {volume} {130}},\
  \bibinfo {pages} {198401} (\bibinfo {year} {2023})}\BibitemShut {NoStop}%
\bibitem [{\citenamefont {Andren\v{s}ek}\ and\ \citenamefont
  {Krajnc}(2025)}]{andrensek25}%
  \BibitemOpen
  \bibfield  {author} {\bibinfo {author} {\bibfnamefont {U.}~\bibnamefont
  {Andren\v{s}ek}}\ and\ \bibinfo {author} {\bibfnamefont {M.}~\bibnamefont
  {Krajnc}},\ }\bibfield  {title} {\bibinfo {title} {Emergent epithelial
  elasticity governed by interfacial surface mechanics and substrate
  interaction},\ }\Eprint {https://arxiv.org/abs/2504.15673} {arXiv:2504.15673}
   (\bibinfo {year} {2025})\BibitemShut {NoStop}%
\bibitem [{\citenamefont {Claussen}\ and\ \citenamefont
  {Brauns}(2025)}]{claussen24}%
  \BibitemOpen
  \bibfield  {author} {\bibinfo {author} {\bibfnamefont {N.~H.}\ \bibnamefont
  {Claussen}}\ and\ \bibinfo {author} {\bibfnamefont {F.}~\bibnamefont
  {Brauns}},\ }\bibfield  {title} {\bibinfo {title} {Mean-field model for
  active plastic flow of epithelial tissue},\ }\href
  {https://doi.org/10.1103/PRXLife.3.023002} {\bibfield  {journal} {\bibinfo
  {journal} {PRX Life}\ }\textbf {\bibinfo {volume} {3}},\ \bibinfo {pages}
  {023002} (\bibinfo {year} {2025})}\BibitemShut {NoStop}%
\bibitem [{\citenamefont {Honda}\ \emph {et~al.}(2004)\citenamefont {Honda},
  \citenamefont {Tanemura},\ and\ \citenamefont {Nagai}}]{honda04}%
  \BibitemOpen
  \bibfield  {author} {\bibinfo {author} {\bibfnamefont {H.}~\bibnamefont
  {Honda}}, \bibinfo {author} {\bibfnamefont {M.}~\bibnamefont {Tanemura}},\
  and\ \bibinfo {author} {\bibfnamefont {T.}~\bibnamefont {Nagai}},\ }\bibfield
   {title} {\bibinfo {title} {A three-dimensional vertex dynamics cell model of
  space-filling polyhedra simulating cell behavior in a cell aggregate},\
  }\href {https://doi.org/10.1016/j.jtbi.2003.10.001} {\bibfield  {journal}
  {\bibinfo  {journal} {J. Theor. Biol.}\ }\textbf {\bibinfo {volume} {226}},\
  \bibinfo {pages} {439} (\bibinfo {year} {2004})}\BibitemShut {NoStop}%
\bibitem [{\citenamefont {Sahu}\ \emph {et~al.}(2021)\citenamefont {Sahu},
  \citenamefont {Schwarz},\ and\ \citenamefont {Manning}}]{sahu21}%
  \BibitemOpen
  \bibfield  {author} {\bibinfo {author} {\bibfnamefont {P.}~\bibnamefont
  {Sahu}}, \bibinfo {author} {\bibfnamefont {J.~M.}\ \bibnamefont {Schwarz}},\
  and\ \bibinfo {author} {\bibfnamefont {M.~L.}\ \bibnamefont {Manning}},\
  }\bibfield  {title} {\bibinfo {title} {Geometric signatures of tissue surface
  tension in a three-dimensional model of confluent tissue},\ }\href
  {https://doi.org/10.1088/1367-2630/ac23f1} {\bibfield  {journal} {\bibinfo
  {journal} {New J. Phys.}\ }\textbf {\bibinfo {volume} {23}},\ \bibinfo
  {pages} {093043} (\bibinfo {year} {2021})}\BibitemShut {NoStop}%
\bibitem [{\citenamefont {Sanematsu}\ \emph {et~al.}(2021)\citenamefont
  {Sanematsu}, \citenamefont {Erdemci-Tandogan}, \citenamefont {Patel},
  \citenamefont {Retzlaff}, \citenamefont {Amack},\ and\ \citenamefont
  {Manning}}]{sanematsu21}%
  \BibitemOpen
  \bibfield  {author} {\bibinfo {author} {\bibfnamefont {P.~C.}\ \bibnamefont
  {Sanematsu}}, \bibinfo {author} {\bibfnamefont {G.}~\bibnamefont
  {Erdemci-Tandogan}}, \bibinfo {author} {\bibfnamefont {H.}~\bibnamefont
  {Patel}}, \bibinfo {author} {\bibfnamefont {E.~M.}\ \bibnamefont {Retzlaff}},
  \bibinfo {author} {\bibfnamefont {J.~D.}\ \bibnamefont {Amack}},\ and\
  \bibinfo {author} {\bibfnamefont {M.~L.}\ \bibnamefont {Manning}},\
  }\bibfield  {title} {\bibinfo {title} {{3D} viscoelastic drag forces
  contribute to cell shape changes during organogenesis in the zebrafish
  embryo},\ }\href {https://doi.org/10.1016/j.cdev.2021.203718} {\bibfield
  {journal} {\bibinfo  {journal} {Cells Dev.}\ }\textbf {\bibinfo {volume}
  {168}},\ \bibinfo {pages} {203718} (\bibinfo {year} {2021})}\BibitemShut
  {NoStop}%
\bibitem [{\citenamefont {Zhang}\ and\ \citenamefont
  {Schwarz}(2022)}]{zhang22}%
  \BibitemOpen
  \bibfield  {author} {\bibinfo {author} {\bibfnamefont {T.}~\bibnamefont
  {Zhang}}\ and\ \bibinfo {author} {\bibfnamefont {J.~M.}\ \bibnamefont
  {Schwarz}},\ }\bibfield  {title} {\bibinfo {title} {Topologically-protected
  interior for three-dimensional confluent cellular collectives},\ }\href
  {https://doi.org/10.1103/PhysRevResearch.4.043148} {\bibfield  {journal}
  {\bibinfo  {journal} {Phys. Rev. Res.}\ }\textbf {\bibinfo {volume} {4}},\
  \bibinfo {pages} {043148} (\bibinfo {year} {2022})}\BibitemShut {NoStop}%
\bibitem [{\citenamefont {Villeneuve}\ \emph {et~al.}(2024)\citenamefont
  {Villeneuve}, \citenamefont {Hashmi}, \citenamefont {Ylivinkka},
  \citenamefont {Lawson-Keister}, \citenamefont {Miroshnikova}, \citenamefont
  {P{\'e}rez-Gonz{\'a}lez}, \citenamefont {Myllym{\"a}ki}, \citenamefont
  {Bertillot}, \citenamefont {Yadav}, \citenamefont {Zhang}, \citenamefont
  {Matic~Vignjevic}, \citenamefont {Mikkola}, \citenamefont {Manning},\ and\
  \citenamefont {Wickstr{\"o}m}}]{villeneuve24}%
  \BibitemOpen
  \bibfield  {author} {\bibinfo {author} {\bibfnamefont {C.}~\bibnamefont
  {Villeneuve}}, \bibinfo {author} {\bibfnamefont {A.}~\bibnamefont {Hashmi}},
  \bibinfo {author} {\bibfnamefont {I.}~\bibnamefont {Ylivinkka}}, \bibinfo
  {author} {\bibfnamefont {E.}~\bibnamefont {Lawson-Keister}}, \bibinfo
  {author} {\bibfnamefont {Y.~A.}\ \bibnamefont {Miroshnikova}}, \bibinfo
  {author} {\bibfnamefont {C.}~\bibnamefont {P{\'e}rez-Gonz{\'a}lez}}, \bibinfo
  {author} {\bibfnamefont {S.-M.}\ \bibnamefont {Myllym{\"a}ki}}, \bibinfo
  {author} {\bibfnamefont {F.}~\bibnamefont {Bertillot}}, \bibinfo {author}
  {\bibfnamefont {B.}~\bibnamefont {Yadav}}, \bibinfo {author} {\bibfnamefont
  {T.}~\bibnamefont {Zhang}}, \bibinfo {author} {\bibfnamefont
  {D.}~\bibnamefont {Matic~Vignjevic}}, \bibinfo {author} {\bibfnamefont
  {M.~L.}\ \bibnamefont {Mikkola}}, \bibinfo {author} {\bibfnamefont {M.~L.}\
  \bibnamefont {Manning}},\ and\ \bibinfo {author} {\bibfnamefont {S.~A.}\
  \bibnamefont {Wickstr{\"o}m}},\ }\bibfield  {title} {\bibinfo {title}
  {Mechanical forces across compartments coordinate cell shape and fate
  transitions to generate tissue architecture},\ }\href
  {https://doi.org/10.1038/s41556-023-01332-4} {\bibfield  {journal} {\bibinfo
  {journal} {Nat. Cell Biol.}\ }\textbf {\bibinfo {volume} {26}},\ \bibinfo
  {pages} {207} (\bibinfo {year} {2024})}\BibitemShut {NoStop}%
\end{thebibliography}%
\end{document}


\renewcommand{\theequation}{S\arabic{equation}}
\renewcommand{\thefigure}{S\arabic{figure}}
\renewcommand{\thepage}{S\arabic{page}}
\renewcommand{\thetable}{S}
\title{Geometry of T1 transitions in epithelia\\\textbullet{} Supplemental Material \textbullet{}}

\author{Pierre A. Haas}
\affiliation{Max Planck Institute for the Physics of Complex Systems, N\"othnitzer Stra\ss e 38, 01187 Dresden, Germany}
\affiliation{\smash{Max Planck Institute of Molecular Cell Biology and Genetics, Pfotenhauerstra\ss e 108, 01307 Dresden, Germany}}
\affiliation{Center for Systems Biology Dresden, Pfotenhauerstra\ss e 108, 01307 Dresden, Germany}\maketitle

In this Supplemental Material, I present numerical results for a mean-field vertex model of an isolated T1 transition that complement the exact results in the main text.

\section*{Mean-field vertex model}
\subsection{Model}
In standard incarnations of the vertex model~\cite{farhadifar07,staple10,fletcher14,fletcher16,alt17}, the nondimensional energy of a single cell of area $A$ and perimeter~$P$ is
\begin{align}
E_\text{vertex}=K(A-\mathcal{A}_0)^2+(P-\mathcal{P}_0)^2,
\end{align}
where $\mathcal{A}_0=1$ is the preferred cell area, $\mathcal{P}_0$ is the preferred cell perimeter, and $K$ is the ratio of the nondimensional area elasticity and perimeter elasticity moduli.
\begin{figure}[b]
\includegraphics{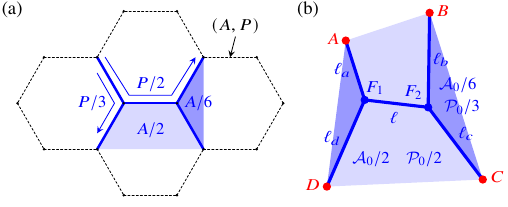}
\caption{Mean-field vertex model of an isolated T1 transition. (a)~Regular lattice of hexagonal cells of area $A$ and perimeter $P$. An edge and its neighbors define two pairs of paths of lengths $P/2$ and $P/3$ respectively (arrows) and two pairs of trapezoids and triangles of respective areas $A/2$ and $A/6$ (shaded regions). (b)~Mean-field description of a T1 transition in a vertex model with preferred area $\mathcal{A}_0=1$ and preferred perimeter $\mathcal{P}_0$: The outer vertices $A,B,C,D$ (red) are fixed as the central edge formed by the inner vertices $F_1,F_2$ (blue) undergoes a T1 transition. The segments joining these points have lengths $\ell_a,\ell_b,\ell_c,\ell_d,\ell$, and define two trapezoidal regions with preferred area $\mathcal{A}_0/2$ and preferred length $\mathcal{P}_0/2$, as well as two triangular regions with preferred area $\mathcal{A}_0/6$ and preferred length $\mathcal{P}_0/3$.\label{figS1}}
\end{figure}

To obtain a mean-field description of an isolated T1 transition in this model, I note that, in a regular hexagonal lattice~\figref{figS1}{a}, a T1 configuration of five edges defines two pairs of paths of lengths $P/3$ and $P/2$ that border two triangular regions of area $A/6$, and two trapezoidal regions of area $A/2$. In my mean-field description of a T1 transition with fixed outer vertices $A,B,C,D$ and inner vertices $F_1,F_2$~\figref{figS1}{b}, I therefore assign to the two triangular regions $AF_1D,BF_2C$ a preferred area $\mathcal{A}_0/6$ and a preferred length $\mathcal{P}_0/3$, and to the two trapezoidal regions $AF_1F_2B,CF_2F_1D$ a preferred area $\mathcal{A}_0/2$ and a preferred length $\mathcal{P}_0/2$. The mean-field energy is thus
\begin{align}
E&=K\left[\left([AF_1F_2B]-\dfrac{\mathcal{A}_0}{2}\right)^2+\left([CF_2F_1D]-\dfrac{\mathcal{A}_0}{2}\right)^2\right.\nonumber\\
&\qquad+\left.\left([BF_2C]-\dfrac{\mathcal{A}_0}{6}\right)^2+\left([DF_1A]-\dfrac{\mathcal{A}_0}{6}\right)^2\right]\nonumber\\
&\quad+\left(\overline{AF_1F_2B}-\dfrac{\mathcal{P}_0}{2}\right)^2+\left(\overline{CF_2F_1D}-\dfrac{\mathcal{P}_0}{2}\right)^2\nonumber\\
&\qquad+\left(\overline{BF_2C}-\dfrac{\mathcal{P}_0}{3}\right)^2+\left(\overline{DF_1A}-\dfrac{\mathcal{P}_0}{3}\right)^2,\label{eq:mf}
\end{align}
in which $[AF_1F_2B],[CF_2F_1D],[BF_2C],[DF_1A]$ denote the respective areas of trapezoids $AF_1F_2B,CF_2F_1D$ and triangles $BCF_2,DAF_1$, $\smash{\overline{AF_1F_2B}}=\ell_a+\ell+\ell_b$, $\smash{\overline{CF_2F_1D}}=\ell_c+\ell+\ell_d$, $\smash{\overline{BF_2C}}=\ell_b+\ell_c$, and $\smash{\overline{DF_1A}}=\ell_d+\ell_a$.

In the units in which $\mathcal{A}_0=1$ that I will use in what follows, it is easy to see that $E=0$ for a regular hexagonal lattice if and only if $\mathcal{P}_0=\mathcal{P}_6\equiv8^{1/2}3^{1/4}\approx 3.72$ and $ABCD$ has area $\mathcal{A}=\mathcal{A}_6\equiv4/3$. These values therefore define natural parameter scales for my numerical calculations.

\begin{figure*}
\centering\includegraphics{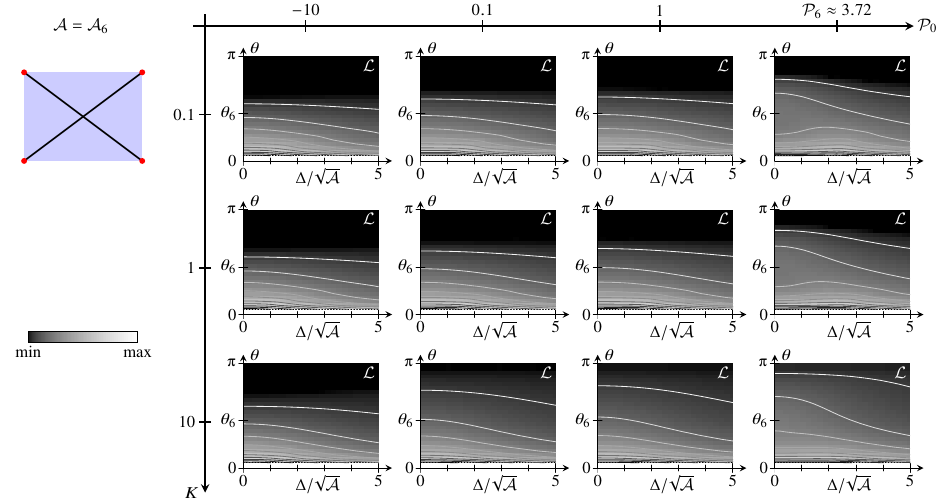}
\caption{Results for the mean-field vertex model. Contour plots of the distance $\mathcal{L}$ from the T1 transition in $(\upDelta,\theta)$ space, for different values of $K$ and $\mathcal{P}_0$ and for fixed $\mathcal{A}=\mathcal{A}_6$ and quadrilaterals with diagonals intersecting at their midpoints (i.e., parallelograms, as shown in the inset). The contour plots show that $\mathcal{L}$ decreases with increasing disorder.}\label{figS2}
\end{figure*}

\begin{figure*}
\centering\includegraphics{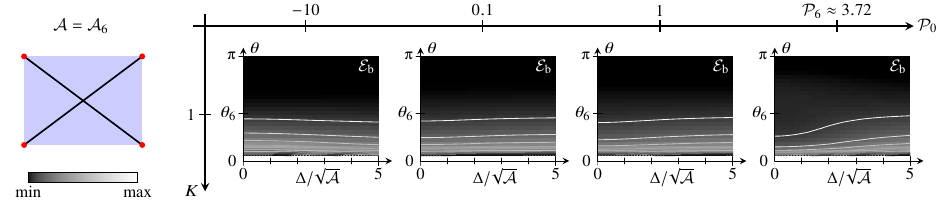}
\caption{Results for the mean-field vertex model. Contour plots of the energy barrier $\mathcal{E}_\text{b}$ to the T1 transition, analogous to \textwholefigref{figS2}, shown for the single value $K=1$. The contour plots show that $\mathcal{L}$ decreases with increasing disorder for $\mathcal{P}_0>0$.}\label{figS3}
\end{figure*}

\subsubsection*{\textbf{Limiting cases}}
Two limiting cases of Eq.~\eqref{eq:mf} are of interest. First, if $K\ll 1$ and $\mathcal{P}_0<0$, $|\mathcal{P}_0|\gg 1$, and all other parameters associated with the geometry of $ABCD$ are of order $O(1)$, then
\begin{align}
E&\approx |\mathcal{P}_0|\left[\overline{AF_1F_2B}+\overline{CF_2F_1B}+\dfrac{2}{3}\left(\overline{BF_2C}+\overline{DF_1A}\right)\right]+\text{const.}\nonumber\\
&\qquad =\dfrac{5|\mathcal{P}_0|}{3}\left(\ell_a+\ell_b+\ell_c+\ell_d+\dfrac{6}{5}\ell\right)+\text{const.}.
\end{align}
Thus, for $K\ll 1$ and $\mathcal{P}_0<0$, $|\mathcal{P}_0|\gg 1$, the system is dominated by linear line tensions. They are not uniform, however, unlike the fully analytically solvable case discussed in the main text: the underlying vertex model structure results in a higher effective line tension on the central edge.

The second case of interest is $K\ll 1$, $\mathcal{P}_0\ll 1$, in which Eq.~\eqref{eq:mf} becomes
\begin{align}
E&\approx\overline{AF_1F_2B}^2+\overline{CF_2F_1B}^2+\overline{BF_2C}^2+\overline{DF_1A}^2\nonumber\\
&\hspace{10pt}=2\bigl(\ell_a^2+\ell_b^2+\ell_c^2+\ell_d^2+\ell^2+\ell_a\ell_b+\ell_b\ell_c+\ell_c\ell_d+\ell_d\ell_a\nonumber\\
&\qquad\qquad+\ell\ell_a+\ell\ell_b+\ell\ell_c+\ell\ell_d\bigr).\label{eq:nl}
\end{align}
The system is now dominated by uniform nonlinear line tensions. These differ, however, from the nonlinear line tensions that I have analyzed in the main text, because there are additional terms coupling the different sides. As also mentioned in the main text, this is the reason why the nonlinear line tensions that I have analyzed there were introduced by Ref.~\cite{grossman22}.

\begin{figure*}
\centering\includegraphics{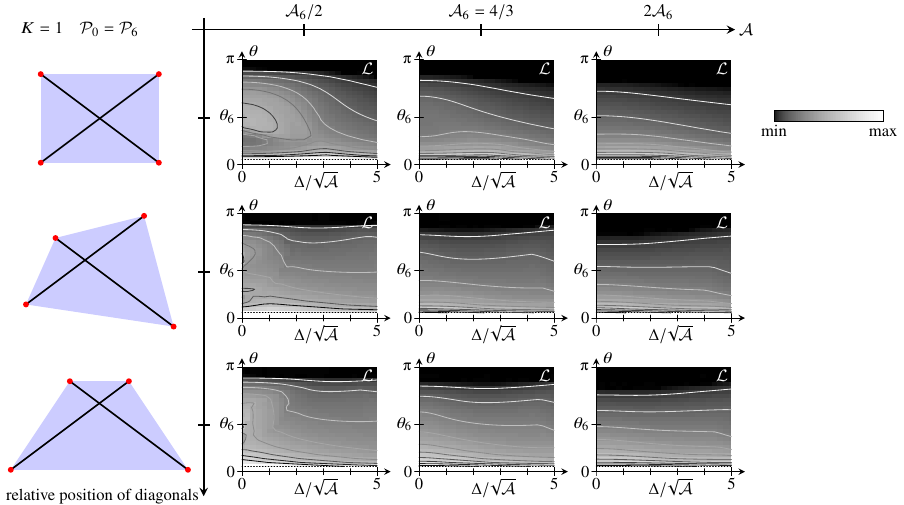}
\caption{Results for the mean-field vertex model. Contour plots of the distance $\mathcal{L}$ from the T1 transition in $(\upDelta,\theta)$ space, for different values of~$\mathcal{A}$ and different relative positions of the intersection of the diagonals of the quadrilateral (inset), and for fixed $K=1$, $\mathcal{P}_6$.}\label{figS4}
\end{figure*}

\subsubsection*{\textbf{Numerical minimization}}
For fixed positions of the outer vertices $A,B,C,D$, I minimize Eq.~\eqref{eq:mf} numerically~\footnote{\textsc{Matlab} code for minimization of this mean-field vertex model energy is available online at \texttt{zenodo.org/...}.} over all positions of the inner vertices $F_1,F_2$ using the \texttt{fmincon} function of \textsc{Matlab} (The MathWorks, Inc.). I impose the constraints that $F_1,F_2$ both lie inside quadrilateral $ABCD$, and that $F_1,F_2$ lie inside triangles $AF_2D,BF_1C$, respectively. The latter conditions ensure that $AF_1F_2B,CF_2F_1D$ are convex. I also minimize $E$ over all positions of a collapsed central vertex $F=F_1=F_2$ inside quadrilateral $ABCD$. These minimizations determine the distance $\mathcal{L}$ from the T1 transition and the energy barrier $\mathcal{E}_\text{b}$ from the transition defined in Fig.~1(c) of the main text.

\vspace{-7pt}
\subsection{Results}
\vspace{-5pt}
I analyze this mean-field vertex model in the equal-area ensemble introduced in Fig.~2(b) of the main text. As I did there, I denote by $\mathcal{D}_1=|AC|$, $\mathcal{D}_2=|BD|$ the diagonals of quadrilateral $ABCD$, and by $\theta$ the angle they make at their intersection~$O$. In the simple models in the main text, I found that, in this equal-area ensemble, $\mathcal{L},\mathcal{E}_\text{b}$ depend on the geometry of $ABCD$ only through $\theta$ and $\upDelta=|\mathcal{D}_1-\mathcal{D}_2|$, but not through the relative positions $(|AO|-|CO|)/\mathcal{D}_1$, $(|BO|-|DO|)/\mathcal{D}_2$ of $O$ along the diagonals. Motivated by these results, I consider families of quadrilaterals of fixed relative positions of the diagonals, and determine $\mathcal{L},\mathcal{E}_\text{b}$ in $(\upDelta,\theta)$ space for different values of the remaining parameters $K,\mathcal{P}_0,\mathcal{A}$.

I plot contours of the distance $\mathcal{L}$ from the T1 transition in~$(\upDelta,\theta)$ space in \textwholefigref{figS2}, for different values of $K$ and $\mathcal{P}_0$, similarly to the plots in Figs.~2(e), 2(f), 3(b), and 3(c) in the main text. These plots show that $\mathcal{L}$ decreases, at constant $\theta$, for large amounts of disorder $\upDelta$. At intermediate values of disorder, more complex behaviour arises for $\mathcal{P}_0\approx\mathcal{P}_6$ \wholefigref{figS2}. In particular, reduced values of $\mathcal{L}$ for $\theta\approx\theta_6$ recall the fluidisation transition of the vertex model~\cite{bi15}, but it is important to recognise that the latter is a nonlocal effect that cannot be captured quantitatively by this local description.

As expected, the results for $K\ll 1$ and $\mathcal{P}_0<0$, $|\mathcal{P}_0|\gg 1$ agree qualitatively with the exact results for (homogeneous) linear line tensions obtained in the main text [Fig.~2(e)]. More surprisingly, the results for $K\ll 1$, $\mathcal{P}_0\ll 1$ do not show the increase of $\mathcal{L}$ with $\upDelta$ obtained in the text for the nonlinear line tensions introduced in Ref.~\cite{grossman22}. This is not a mathematical contradiction, because the latter are different from the nonlinear tensions in Eq.~\eqref{eq:nl} that appear in this limit of my mean-field vertex model.

Interestingly, the energy barrier $\mathcal{E}_\text{b}$ increases with increasing disorder $\upDelta$~\wholefigref{figS3} despite the decrease of $\mathcal{L}$~\wholefigref{figS2}, except for $\mathcal{P}_0<0$, $|\mathcal{P}_0|\gg 1$, in which case the energy barrier decreases with increasing $\upDelta$~\wholefigref{figS3} in agreement with the exact results for linear line tensions in the main text. I stress that this does not contradict the generic scaling $\mathcal{E}_\text{b}\sim\mathcal{L}^2$ expected for small $\mathcal{L}$ from Ref.~\cite{popovic21}, because this difference only arises for sufficiently large $\mathcal{E}_\text{b}$ or $\mathcal{L}$.

Unlike the exact results in the main text, these results do depend on the relative position of the two diagonals of $ABCD$. In \textwholefigref{figS4}, I therefore plot contours of $\mathcal{L}$ for different values of~$\mathcal{A}$ and different relative positions of these diagonals to highlight more complex behaviour. This stresses the effect of cell size and of irregular cell shapes on T1 transitions.

\vspace{16pt}
\section*{Supplemental File}
\vspace{-4pt}
The Supplemental File \texttt{SM.zip} contains \textsc{Mathematica} (Wolfram, Inc.) notebooks with details of the exact calculations in the main text.
\bibliography{main}